\documentclass[twocolumn,usenatbib,useAMS]{mn2e}

\usepackage{natbib}
\usepackage{amsbsy}
\usepackage{amssymb}
\usepackage{amsmath}
\usepackage{graphicx}
\usepackage[caption=false]{subfig}
\usepackage [english]{babel}
\usepackage [english]{babel}
\begin{document}

\title[Interglitch dynamics, braking index of Vela and time to next glitch]{Nonlinear interglitch dynamics, the braking index of the Vela pulsar and the time to the next glitch}

\author[Akbal et al.]{ O. Akbal$^1$\thanks{E-mail:
oakbal@sabanciuniv.edu}, M. A. Alpar$^{1}$, S. Buchner$^{2,3,4}$\& D. Pines$^5$ \\
  \\
  $^1$Sabanc{\i} University, Faculty of Engineering and Natural Sciences, Orhanl{\i}, 34956 Istanbul, Turkey \\
  $^2$Square Kilometer Array South Africa, The Park, Park Road, Pinelands, Cape Town 7405, South Africa\\
  $^3$Hartebeesthoek Radio Astronomy Observatory, P.O. Box 443, Krugersdorp 1740, South Africa\\
  $^4$School of Physics, University of the Witwatersrand, PO BOX Wits, Johannesburg 2050, South Africa\\ 
  $^5$Santa Fe Institute, 1399 Hyde Park Rd., Santa Fe NM 87501, USA 
  }

\maketitle

\begin{abstract}
The inter-glitch timing of the Vela pulsar is characterized by a constant second derivative of the rotation rate. This takes over after the post-glitch exponential relaxation, and is completed at about the time of the next glitch. The vortex creep model explains the second derivatives in terms of non-linear response to the glitch. We present inter-glitch timing fits to the present sample covering 16 large glitches, taking into account the possibility that in some glitches part of the step in spin-down rate may involve a ``persistent shift'', as observed in the Crab pulsar. Modifying the expression for the time between glitches with this hypothesis leads to better agreement with the observed inter-glitch time intervals. We extrapolate the inter-glitch model fits to obtain spin-down rates just prior to each glitch, and use these to calculate the braking index $n = 2.81 \pm 0.12$. The next glitch should occur around Dec. 22, 2017
$\pm  197$  days if no persistent shift is involved, but  could occur as early as July 27, 2016 $\pm 152$ days if the 2013 glitch gave rise to a typical Vela persistent shift.
\textbf{Note added:} Literally while we were submitting the first version of this paper, on Dec. 12, 2016, we saw ATel $\#$ 9847 announcing a Vela pulsar glitch which has arrived 138 days after our prediction with a persistent shift, within the $1 \sigma$ uncertainty of $152$ days.
\end{abstract}


\section{Introduction}
\label{sec:intro}  
The Vela pulsar, PSR 0833-45, was the first pulsar for which a glitch, a sudden speed-up, in spin-down behaviour was observed \citep{radhakrishnan69, reichley69}, and it continues to be one of the most active glitching pulsars, as 15 additional large glitch events have been observed in the succeeding 47 years\footnote{http://www.jb.man.ac.uk/pulsar/glitches/gTable.html}. Because it has been monitored almost daily since 1985, it has proved possible to determine the onset of glitches with an uncertainty of less than a day and to follow in detail its post-glitch behaviour for the 10 glitches that have occurred since that time \citep{mcCulloch87, flanagan91, flanagan94, wang00, yu13, dodson02, dodson04, flanagan06, buchner10, buchner13}. The 2000 glitch, in particular, happened during an observation session, so that an upper limit of $ 40 $ s could be placed on the glitch rise time \citep{dodson02}. In 1994 the Vela pulsar exhibited two large glitches with $ \Delta \Omega/\Omega = 8.61 \times 10^{-7} $ and $ \Delta \Omega/\Omega = 1.99 \times 10^{-7} $ that were separated by just $32$ days  \citep{buchner11}. The glitch rates, signatures and interglitch behaviour of many pulsars, and the statistics of pulsar glitches, indicate similarities to the Vela pulsar \citep{alpar94, alpar06} once one scales with pulsar rotation frequency and spindown rate. We believe  the Vela pulsar is a Rosetta stone for understanding pulsar glitches and the results presented here have general applicability to pulsar dynamics.

It was recognized early on \citep{baym69} that the fact that one is able to see a well-defined glitch relaxation meant the neutrons present in the crust and core of a pulsar must be superfluid. The most successful phenomenological explanation of the origin of pulsar glitches and post-glitch behaviour has been based on the proposals by \cite{packard72} and \cite{anderson75} that glitches are an intrinsic property of the rotating superfluid and originate in the unpinning of vortices that are pinned to crustal nuclei. In this picture, the pinned superfluid will lag the spin-down of the pulsar until a critical angular velocity is reached, at which point the pinned vortices unpin and move rapidly, transferring angular momentum to the crust, which is observed as the glitch. Apart from the sudden unpinning events, vortices also creep, by thermal activation against the pinning barriers, allowing the superfluid to spin down \citep{alpar84a, alpar84b}. 

The vortex creep model posits two different kinds of response to a glitch; a linear response characterized by exponential relaxation, and a non-linear response that is characterized by a step in the spin-down rate and subsequent interglitch recovery with an approximately constant second derivative of the rotation frequency \citep{alpar84b, alpar89}. The model has proven successful in that it provides a natural explanation for the observed transient exponential decays and for the long term recovery extending until the next glitch, which can only be explained by non-linear dynamics. The model also provides a rough estimate of the inter-glitch intervals. 

This understanding of the systematic inter-glitch timing behaviour due to {\em internal torques} between the crust and superfluid components of the neutron star makes it possible to take account of these internal torques to obtain the braking index $n$ that characterizes the {\em external} pulsar torque, and, as we shall see, leads to results that are quite different from an analysis that does not take the internal torques fully into account.

Our aim in the present paper is threefold. First, we study the recovery of the glitch $ \Delta \dot{\Omega}$ in the spindown rate. We confirm that after the initial exponential relaxation components, the negative step $ \Delta \dot{\Omega}$ recovers with an approximately constant second derivative $ \ddot{\Omega} $ of the rotation rate. As observed in the interglitch intervals up to the ninth observed glitch \citep{alpar93} this recovery is completed at about the time of arrival of the next glitch. Indeed, in a multicomponent dynamical system, the response of internal torques to offsets from steady state always leads to eventual recovery of the steady state. The question is whether the next glitch arrives before or after the completion of the recovery after the previous glitch. Motivated by the earlier observation that the recovery was completed at about the time of arrival of the next glitch in the earlier work on 9 glitches, we now investigate if the coincidence between the time of completion of the recovery from one glitch and the time of arrival of the next glitch is still oberved in the present sample of gllitches. This behaviour of complete recovery with constant $ \ddot{\Omega} $ is confirmed in all interglitch intervals in the present sample. We confirm, model independently, that the recovery is completed at about the time of arrival of the next glitch. As any power law behaviour, this recovery reflects some underlying nonlinear dynamics. We next confirm, with the current data set, that the nonlinear creep model continues to fit the data for post-glitch behaviour and enables one to estimate the time intervals between glitches; we find that the accuracy of the estimated times to the next glitch has improved with the doubling of the glitch sample. Secondly, we explore the possibility that some Vela glitches are accompanied by ``persistent shifts'' in the spin-down rate of the kind observed in the Crab pulsar, where a sudden decrease in the spin-down rate at the time of a glitch is observed to remain constant, with no healing, until the arrival of the next glitch, at which it is superseded with an additional similar shift. This hypothesis leads to somewhat shorter estimates of the time to the next glitch, which agree better with the observed time intervals. Thus the persistent shift hypothesis improves the accuracy of the estimates. Third, based on our understanding of the internal torques, and in particular the correlation between the time when the nonlinear creep response is completed and the time of the next glitch, supported by our estimates, we note that the best fiducial epoch for determining the braking index $n$ due to the external (pulsar) torque is the epoch when the response of internal torques to the previous glitch have been complated, i.e. just before the next glitch. Fitting a long term `true' pulsar second derivative of the rotation rate to the spindown rates at these epochs, we find that the Vela pulsar's braking index $n = 2.81 \pm 0.12$, in agreement with most other measured pulsar braking indices, which lie between $ n = 1.8  $ and $n = 3$ for an isolated ideal dipole rotating in vacuum \citep{melatos97, lyne15, antonopoulou15, archibald15, clark16}. In their determination of the Vela pulsar's braking index, \cite{lyne96} assumed the effects of internal torques would be over with the exponential relaxations, and took a fiducial time of 150 days after each glitch to derive a much lower ``anomalous'' braking index $n = 1.4 \pm 0.2$. Recent work \citep{espinoza16} takes into account a term with recovery at a constant $ \ddot{\Omega} $, but assumes that the recovery is not completed. The braking index $ n=1.7 \pm 0.2 $ is obtained. As we show in this paper the recovery at constant interglitch  $ \ddot{\Omega} $ is actually completed just before the next glitch. We find $n = 2.81 \pm 0.12$.

The plan of our paper is as follows. Sec. 2 contains a summary of the vortex creep model that is used, in Sec. 3, to fit all currently available inter-glitch timing data. Sec. 4 describes the estimation of inter-glitch time intervals, with or without the inclusion of persistent shifts. Sec. 5 derives the braking index of the Vela pulsar for several  different variants of our model. Sec. 6 contains our conclusions.

\section{The Vortex Creep Model}
\label{sec:creep}

The postglitch behaviour of the Vela pulsar exhibits both linear response, in the form of three distinct exponential relaxations, and a non-linear response with constant $ \ddot{\Omega}$ that persists until the next glitch. The exponential decays have relaxation times $\tau \lesssim 32$ days. The vortex creep model \citep{alpar84a, alpar84b, alpar89} explains glitches and postglitch behaviour in terms of superfluid dynamics that takes into account vortex pinning, unpinning and creep. We summarize the main concepts here, referring the reader to earlier work for details.

The superfluid components of the neutron star rotate by sustaining quantized vortices, and their spin down in response to the pulsar torque is described by the motion of these vortices radially outward from the rotation axis. Vortex motion is impeded by pinning to inhomogeneities, such as the nuclei in the neutron star's inner crust where the neutron superfluid coexists with the crustal crystalline lattice \citep{alpar77} or to toroidal flux lines in the outer core of the neutron star \citep{guger14}. Vortices unpin and repin by thermal activation, thereby providing a vortex creep current. Because of pinning, the superfluid rotates somewhat faster than the crust as the crust spins down under the external torque. The lag $ \omega \equiv \Omega_{s}-\Omega_{c} $ between the rotation rates of the superfluid and crust provides a bias to drive a vortex creep current in the radially outward direction from the rotation axis. Vortex creep thus allows the superfluid to spin down. This process has a steady state when both superfluid and normal matter are spinning down at the same rate, driven by a steady state lag $ \omega_{\infty} $.

If the lag reaches the maximum value $ \omega_{cr} > \omega_{\infty}$ that can be sustained by the pinning forces, vortices unpin and move outward rapidly in an avalanche, thereby transferring angular momentum to the crust, leading to the glitch \citep{anderson75, packard72}. In addition to the parts of the pinned superfluid with continuous vortex current, analogous to resistors in an electric circuit, there are also vortex traps prone to catastrophic unpinning, interspersed with vortex free regions, analogous to capacitors. Vortex traps are sites of extra pinning strength, where critical conditions for unpinning can be reached due to this enhanced vortex density \citep{chauetal93, mochizuki95}. The high vortex density in the traps leads to local superfluid velocities that are too large to permit pinned vortices in the regions surrounding the traps, which are therefore vortex free regions, containing few and ineffective pinning centres. Vortex free regions contribute to the angular momentum transfer at glitches, but do not contribute to the spin-down between glitches since they do not sustain vortices participating in the creep process. 

When critical conditions $ \omega = \omega_{cr}$ are reached, vortices unpin collectively from vortex traps, scatter through the vortex free regions, and incite unpinning at further vortex traps, and so create an avalanche. This sudden transfer of angular momentum to the crust, analogous to charge transfer in capacitor discharges, is  observed as a  glitch in the steady spin-down of the pulsar. Glitches by critical unpinning superposed on an underlying vortex creep process were simulated by a ``coherent noise" statistical model by \citet{MelatosWarszawski09} who derive interesting information on pinning parameters by comparison of their simulations with the statistics of frequently glitching pulsars. Further interesting simulations were made by \citet{WarszawskiMelatos11} in a model where glitches by vortex unpinning are  superposed on a continuous relaxation of the superfluid described by the Gross-Pitaevskii Equation. The constraints obtained in these papers on pinning, critical conditions for unpinning and vortex density distributions, with traps and vortex free regions are in agreement with the framework of vortex creep theory.

The sudden increase in the crust rotation rate $\Omega_{c}$ and the decrease in the superfluid rotation rate $\Omega_{s}$ at a glitch offset the lag $\omega$ from its pre-glitch steady state value. If the creep process has a linear dependence on the lag, the response to the offset is simple exponential relaxation. Several components of exponential relaxation are observed in the Vela pulsar. After the exponential relaxation is over the relaxation of the glitch continues, actually until the time of the next glitch, in a characteristic non-exponential manner. The glitch in the spindown rate recovers approximately linearly in time, i.e., to lowest order, with a constant second derivative $\ddot{\Omega}$ of the rotation rate. Just as exponential relaxation is the signature of linear dynamics, power law relaxation processes indicate nonlinear dynamics. It is an appealing feature of the vortex creep model that both types of observed post-glitch behaviour can occur, in different parts of the neutron star superfluid, as two regimes of the same physical process. Being a process of thermal activation, the creep process has an intrinsic exponential dependence on pinning energies, temperature and the driving lag $\omega$ through Boltzmann factors. Depending on the pinning energy $ E_{p} $, the temperature, and the steady state spin-down rate $ \dot{\Omega}_{\infty} $ dictated by the pulsar torque, the creep current in some parts of the superfluid can have a linear dependence on the lag, leading to exponential postglitch response; while other parts of the superfluid have the full nonlinear dependence on the glitch induced perturbation \citep{alpar89}. In the nonlinear regime the steady state lag is very close to the critical lag, $ \omega_{cr}-\omega_{\infty} \ll \omega_{cr} $. This makes it possible to reach the critical conditions for unpinning by fluctuations from steady state as supported by simulations \citep{MelatosWarszawski09}.

The focus of this paper is on the nonlinear creep response by which we model the interglitch recovery at constant $\ddot{\Omega}$, after the exponential relaxation is over. When the superfluid rotation rate is reduced by $\delta\Omega$ as vortices unpinned at the glitch move through a nonlinear creep region of moment of inertia $\delta I$, the non-linear creep current, with its very sensitive dependence on the lag, will stop. This region is not spinning down after the glitch. As the pulsar torque is acting on less moment of inertia, the spindown rate will increase by a step $\Delta\dot{\Omega}$, such that $\Delta\dot{\Omega}/\dot{\Omega} = \delta I/I$ where $I$ is the total moment of inertia of the star. The lag will return to its pre-glitch value after a waiting time $t_{0} \equiv \delta\Omega/|\dot{\Omega}|$, 
since the star continues to spin down under the pulsar torque. Around time $t_{0}$ the creep will start again, showing up as a positive step recovery $|\Delta\dot{\Omega}| = |\dot{\Omega}| \delta I/I$, which will be observable if $\delta I$ is large enough. This extreme nonlinear signature of stop-hold-and-restart was a prediction of the vortex creep model, dubbed ``Fermi function behaviour'' \citep{alpar84a}. This has been observed clearly in one instance in the Vela pulsar \citep{flanagan95,buchner08}. Figure 1 shows a sketch of the predicted behaviour, and the observed step recovery for 400 days after the 1988 glitch of the Vela pulsar, providing strong direct evidence for the presence of nonlinear creep.  

\begin{figure}
\centering

\subfloat[]{\includegraphics[width = 3in,height=4.5cm]{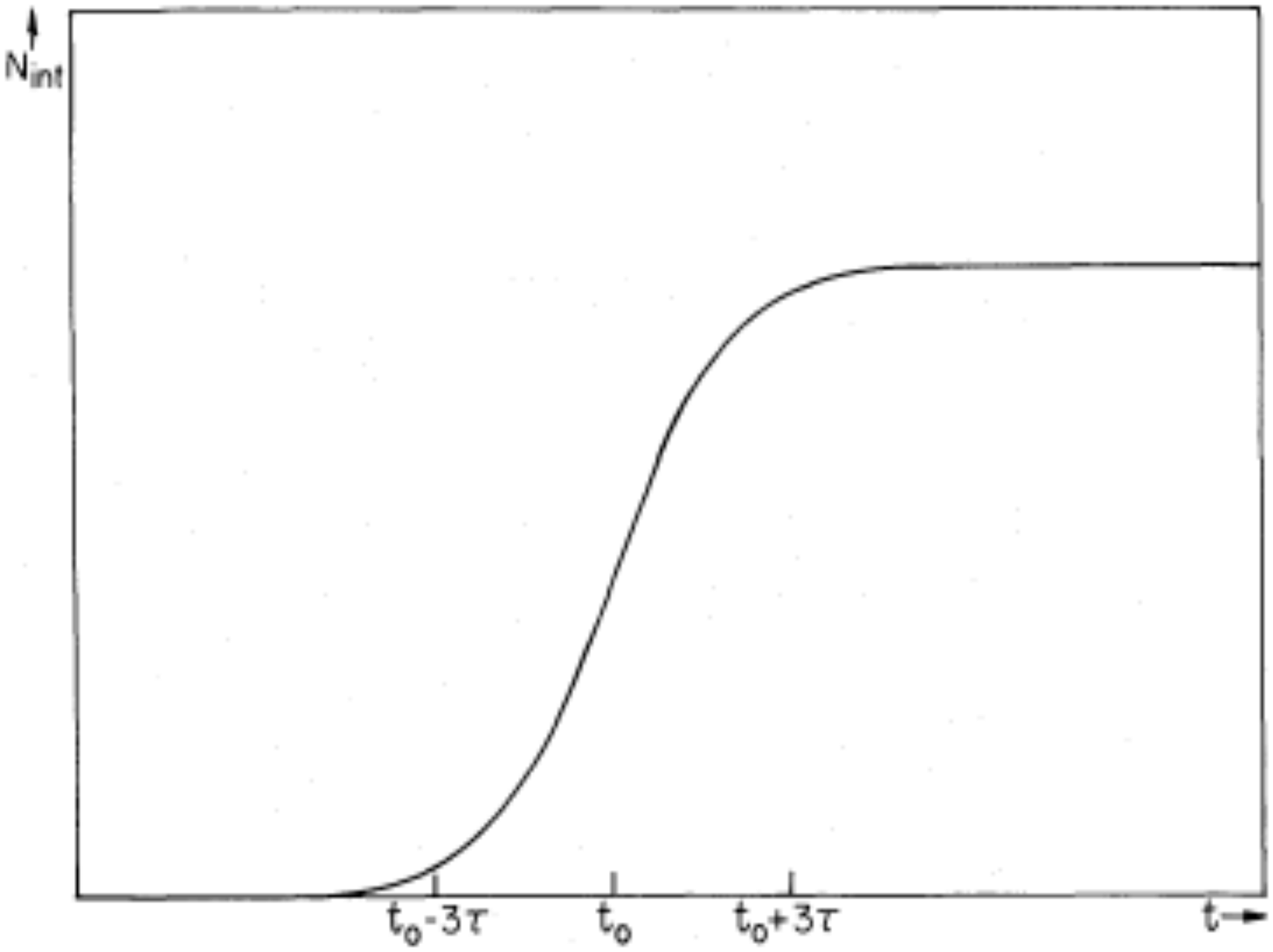}} \\
\subfloat[]{\includegraphics[width = 3in,height=4.5cm]{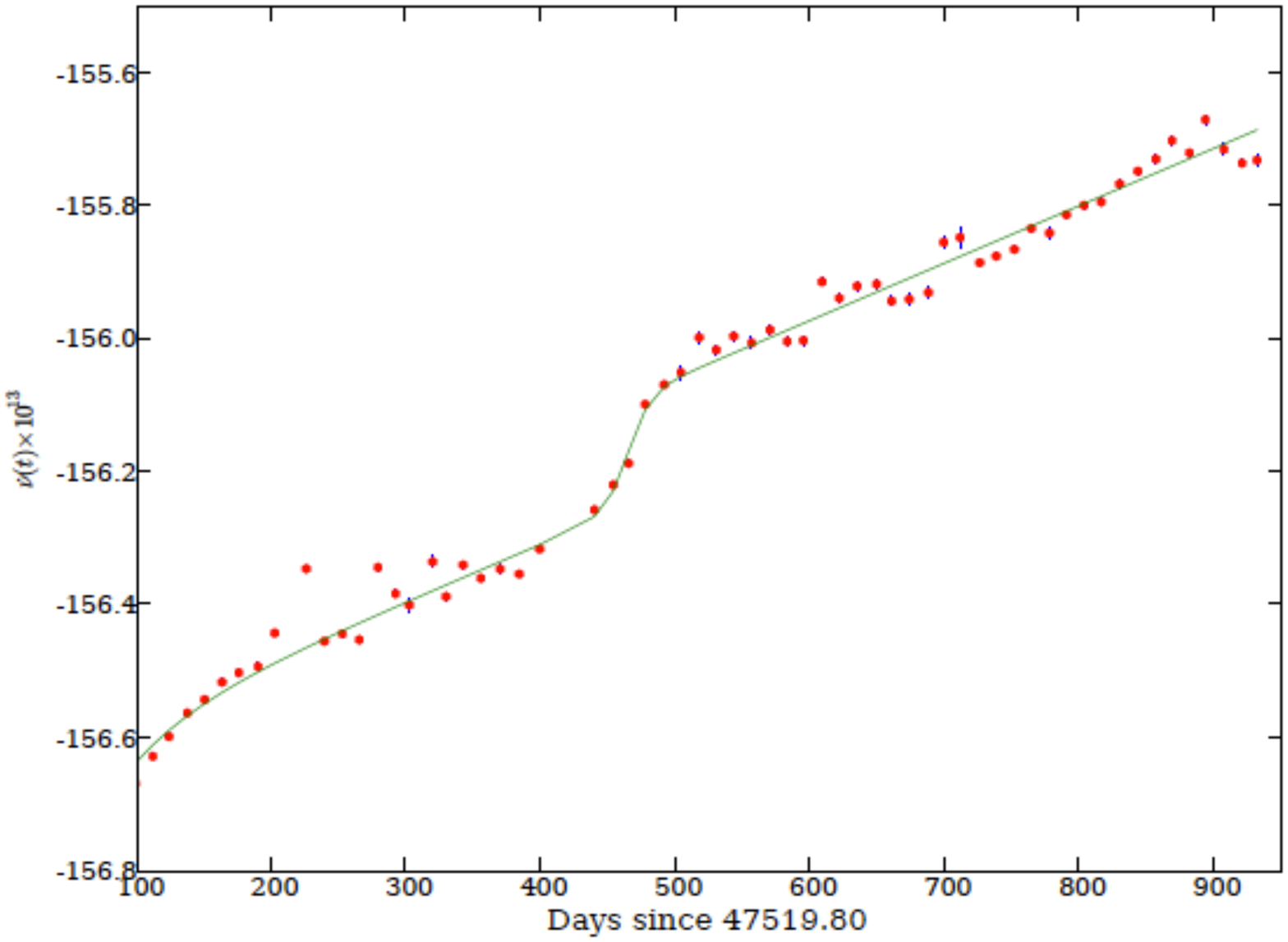}}
\caption{(a) Predicted ``Fermi function behaviour''. Part of the glitch in the spindown rate, $ \Delta \dot{\Omega}=(\delta I/I)\dot{\Omega} $, recovers at time $ t_{0}=\delta \Omega / |\dot{\Omega}| $. This figure is taken from \citet{alpar84a}.  (b) The observed step recovery, at $ t_{0} \cong 400 $ days after the 1988 glitch, is shown, superposed on the ``triangle recovery''. This figure is taken from \citet{buchner08}.}
\label{step}
\end{figure}

More likely, the vortices unpinned at the glitch will cause the superfluid rotation rate to decrease by amounts $\delta\Omega_i$ as they move through many nonlinear creep regions $i$ with moments of inertia $\delta I_i$. Note that $\delta I_i$ does not refer to an actual increase or decrease in the moment of inertia; it denotes a {\em small} amount of moment of inertia belonging to the creep region $i$ where the superfluid rotation rate has decreased at the glitch by the particular amount $\delta\Omega_i$. At the time of the glitch creep stops in all of these regions, causing a total increase in the spindown rate by
\begin{equation}
\frac{\Delta \dot \Omega(0)}{\dot \Omega}=\frac{\Sigma_i \delta I_i}{I} \equiv \frac{I_{A}}{I},
\label{doto}
\end{equation}  
where $I_{A}$ is the total moment of inertia of nonlinear creep regions through which vortices moved suddenly at the glitch. Each component of the inertial moment, $\delta I_i$, will restart creep at its own waiting time $t_i \equiv \delta\Omega_i/|\dot{\Omega}|$; these are not discernible as step recovery in the spin-down rate, as the individual $\delta I_i$ are too small. Instead a stacked response of the nonlinear recovery of creep regions with moments of inertia $\delta I_i$ at successive times $t_i$ will be observed. In particular, in the lowest, mean field approximation, if a uniform average area density of vortices unpins, moves through nonlinear creep regions of total moment of inertia $I_{A}$, and repins, the cumulative response is a ``triangle recovery'' of the glitch step $\Delta \dot \Omega$, with a constant second derivative of the rotation rate
\begin{equation}
\ddot{\Omega} = \frac{I_{A}}{I} \frac{{\dot\Omega}^2}{\delta\Omega} 
\label{ddoto}
\end{equation}
where $\delta\Omega$ is the maximum glitch induced decrease in the superfluid rotation rate, corresponding to the total number of vortices involved in the glitch.  This ``triangle recovery'', as a lowest order approximate fit to the data has been observed to extend till the next glitch in every inter- glitch interval of the Vela pulsar. The end of the triangle response signals the return of the average vortex density to pre-glitch conditions. Assuming that glitches start at critical conditions involving vortex density, the recovery time 
\begin{equation}
t_g \equiv \frac{\delta\Omega}{|\dot{\Omega}|} \equiv \frac{|\Delta\dot{\Omega}|}{\ddot{\Omega}} 
\end{equation}
is roughly when the star is ready for the next glitch. Thus $t_g$ provides a first estimate of the time to the next glitch. The triangle recovery can be written as 
\begin{equation}
\frac{\Delta \dot{\Omega}(t)}{\dot{\Omega}} = \frac{I_{A}}{I} \left( 1 - \frac{t}{t_{g}} \right). 
\end{equation}
To the extent that the density of unpinned vortices is not uniform, the observed post-glitch $\dot{\Omega}(t)$ will deviate from a linear time dependence (constant $\ddot{\Omega}$). The $t^2$ and higher order polynomial terms in $\dot{\Omega}(t)$ are small compared to the lowest order, constant $\ddot{\Omega}$ term. In our model this reflects the dominance of the uniform term in the spatial distribution of the unpinned vortex density $\delta n(r)\equiv \delta n_0$, higher moments (spatial fluctuations) of $\delta n(r)$ being weaker than the mean $\delta n_0$.

Equations (\ref{doto}) and (\ref{ddoto}) are complemented by the angular momentum balance at the glitch when the motion of unpinned vortices leads to a reduction in superfluid rotation rate with maximum value $\delta\Omega$. The glitch induced decrease in the superfluid rotation rate, $\delta\Omega_i$, varies linearly with $\delta I_i$, between $0$ and the maximum value $\delta\Omega$, throughout the constituent small regions with moments of inertia $\delta I_i$ that collectively make up the creep regions with total moment of inertia $I_{A}$. The average is $\delta\Omega/2$. The total angular momentum 
transfer in the glitch from the nonlinear creep regions in the superfluid to the crust is therefore $(1/2) I_{A}  \delta\Omega$. This angular momentum is transferred back to the superfluid during the triangle recovery. The vortices unpinned at the glitch also move through the vortex free (capacitor) regions, of total moment of inertia $I_{B}$, interspersed with the nonlinear creep regions, yielding an additional angular momentum transfer $I_{B}\delta\Omega$, which is not returned to the superfluid. This irreversible vortex discharge at the glitch is analogous to capacitor discharges. It accounts for the part of the glitch in frequency that does not relax back, as is seen in the observations. The total angular momentum lost by these components of the superfluid accounts for the observed spin-up $\Delta \Omega_c$ that remains after the exponential relaxations:
\begin{equation}
I_{c}\Delta \Omega_c = \left( \frac{1}{2}I_{A} + I_{B} \right) \delta \Omega.
\end{equation}

Equations (1), (2) and (5) constitute the nonlinear vortex creep model relating the long term interglitch behaviour to the glitch parameters. Using the observed values of the glitch in rotation frequency, $\Delta \Omega_c$, and in spin down rate, $\Delta \dot{\Omega}$, after the exponential relaxations are subtracted, together with the observed interglitch $\ddot{\Omega}$, these three equations can be solved for the model parameters $ I_{A} $, $ I_{B} $, and $ \delta \Omega$, or, equivalently, for $t_{g}$, the estimated time to the next glitch. We shall call this our Model (1). Note that the interglitch recovery depicted in Eq. (4) is by itself a model independent phenomenological description of the observed timing behaviour. It can be used as a basis for braking index calculations that take into account the full interglitch behaviour extending till the next glitch. The vortex creep model provides the physical context for understanding the power law (constant $\ddot{\Omega}$) interglitch timing which is a clear signal of  nonlinear dynamical behaviour. Moreover,  the natural relation between the glitch parameters provided by the vortex creep model, has been found to have remarkable predictive power for the time to the next glitch for the majority of glitch intervals, while  the estimates can be improved by  extending  the vortex creep model.

Shannon et al (2016) have modeled the interglitch timing with exponential recovery with a constant relaxation time $\tau = 1600$ days, and ascribe the residuals to a noise process with a power law spectrum. They find that each interglitch interval and the total spindown rate data set have similar power spectra and conclude that the noise process is stationary. Their calculated  spindown rate time series however does not contain the frequency jumps at glitches, an essential ingredient in the vortex creep model for the determination of the subsequent second derivative $\ddot{\Omega}$ of the rotation rate. The fact that $\ddot{\Omega}$ correlates with the glitch parameters in accordance with Eqs. (1), (2) and (5) therefore favours the vortex creep model versus a noise process that can reproduce the same behaviour. Employing the standard formula for the braking index, $n \equiv \Omega\ddot{\Omega}/\dot{\Omega}^2$ to the constant $\ddot{\Omega}$ due to the internal torques leads to an ``anomalous'' inter-glitch braking index given by Eqs. (1), (2) and (5); 
\begin{equation}
n_{ig} \equiv \frac{\Omega \ddot{\Omega}_{int}}{\dot{\Omega}^{2}}=\left(\frac{1}{2}+ \frac{I_{A}}{I_{B}} \right)\left( \frac{\Delta \dot{\Omega}}{\dot{\Omega}} \right)^{2}_{-3} \left( \frac{\Delta \Omega}{\Omega} \right)_{-6}^{-1}.  
\end{equation}  
The subscripts -3 and -6 mean the corresponding dimensionless ratios are normalized to 10$^{-3}$ and 10$^{-6}$ respectively.

\section{Observations and Model Fitting} \label{sec:fit}
Since 1985 the Vela pulsar has been observed on most days using the Hartebeesthoek Radio Astronomy Observatory (HartRAO) telescope. Observations are made at either 1668 MHz or 2273 MHz using the 26 m telescope. The South polar bearing failed on 3 Oct 2008 and was repaired on 22 July 2010. During this period the telescope could only be driven in declination. Observations were made of Vela as it transited. The MeerKAT precursor XDM (now HartRAO 15m) was used for Vela observations from 6 Feb 2009 until the telescope was repaired. The Tempo2 glitch plugin \citep{edwards06, hobbs06} was used to calculate $ \dot{\Omega} $ values. 

\cite{alpar84b} initially analysed the postglitch behaviour of the Vela pulsar for its first four glitches within the vortex creep model. This work used data with major uncertainties in the actual dates of the glitches. \cite{alpar93} and \cite{chauetal93} then examined the post-glitch recovery of the first nine glitches of Vela pulsar. They used the vortex creep model in which the post-glitch relaxation is described by the equation: 
\begin{align}
\frac{\Delta \dot{\Omega}_{c}(t)}{|\dot{\Omega}|_{\infty}}= -\sum_{i=1}^{3} \frac{I_{i}}{I} \frac{\Delta \Omega_{c} (0)}{|\dot{\Omega}|_{\infty} \tau_{i}} e^{-t/ \tau_{i}} -\frac{I_{A}}{I}+\frac{I_{A}}{It_{g}}t.
\end{align}
The first three terms express the short-term exponential relaxation response to a glitch. The remaining two terms are relevant for the long time scale and describe the nonlinear response. Exponential relaxation components with timescales of hours, days and months are commonly observed after Vela pulsar glitches. Theoretical interpretations of the exponential relaxation terms include superfluid mutual friction (eg \citet{haskell14}), vortex creep in the linear regime \cite{alpar89}, vortex - lattice interactions \cite{jones93} and Ekman pumping \cite{vanEysden10}. \cite{alpar93} found that $ \tau_{1}=10  $ hr, $ \tau_{2}= 3.2$ days, and $ \tau_{3}=32 $ days, described  all Vela glitches. The exponential relaxations were followed by the long term triangle recovery of the spindown rate, from which they extracted estimates of the time $t_g$ to the next glitch.

We follow the fitting procedure of the earlier applications to analyze the long term relaxation of the 1994 double glitch and the 1996, 2000, 2004, 2006, 2010 and 2013 glitches. For these later glitches we use data from the Hartebesthoek Observatory. We use the Levendberg-Marquardt method to find the best fit values with MPFITFUN procedure \citep{markwardt09}\footnote{ http://purl.com/net/mpfit}. A quick look at the data shows that by 100 days after each glitch all exponential relaxation components are fully relaxed \citep{yu13}. We therefore use data starting from 100 days after each glitch for our long term interglitch fits with the last two terms in Equation (7) which describe the ``triangle'' nonlinear creep response, Eq. (4). The fits are shown in Figure 2.
 
The best fitting constant $\ddot{\Omega}$ values, inferred parameters $ I_{A}/I $, $ I_{B}/I $  and $ t_{g} $ and observed times $ t_{obs} $ to the next glitch are tabulated for the current sample of 17 glitches in Table 1 which presents the results of \cite{alpar93} for the first 8 glitches, of \cite{chauetal93} for the 9th glitch, and our results for the last 8 glitches. Errors quoted for $\ddot{\Omega}$ are formal linear regression errors, which propagate to give the same percentage errors in $n_{ig}$. For the most recent, 2013, glitch we have an estimated time to the next glitch, $t_g = 1553$ d, which gives the expected glitch date as Dec. 22, 2017. The moment of inertia fractions $ I_{1}/I $, $ I_{2}/I $ and $ I_{3}/I $ associated with exponential relaxations with $ \tau_{1}=10 $ hr, $ \tau_{2}=3.2 $ days and $ \tau_{3}=32 $ days are also tabulated. These are not relevant to the long term recovery discussed in this paper. They do, however, contribute to lower bounds on the moment of inertia of creeping superfluid, which in turn are relevant to possible constraints on the equation of state of neutron star matter \citep{datta93, link99}. $ I_{1}/I $, $ I_{2}/I $ and $ I_{3}/I $ are included in Table 1.

As a measure of the estimates  we shall use
\begin{equation}
\Delta t_{i} \equiv t_{g,i}-t_{obs,i}. 
\end{equation}
For the first 9 glitches, analyzed earlier, the mean $\overline{\Delta t}(9) = 200  $ days and the standard deviation $\sigma (9) = 321 $ days. Interestingly, for 7 out of the first 9 glitches the estimator $t_{g}$ was longer than $ t_{obs} $. For the 
current sample the mean fractional deviation between the estimated and observed intervals between the glitches is now $\overline{\Delta t}(15) = 132  $ days and the standard deviation $\sigma(15)= 256 $ days for the 15 inter-glitch intervals. For 11 out of the 15 glitches with observed times to the next glitch, the estimated $ t_{g}$ is longer than $ t_{obs} $.  As the sample of glitches with an observed time to the next glitch has increased from 9 to 15, the mean accuracy of the estimate has decreased from $\overline{\Delta t}(9) = 200  $ days to $\overline{\Delta t}(15) = 132  $ days, with the standard deviation decreasing from $\sigma (9) = 321 $ days to $\sigma(15)= 256 $ days. This strongly supports the validity of the nonlinear vortex creep model, in particular to the association of the interglitch $\ddot{\Omega}$ with a recovery process the completion of which indicates the re-establishment conditions prone to a glitch. It is significant that the mean offset of our estimates $t_g$ from the observed glitch intervals $t_{obs} $ remains positive, even though the sample almost doubled. Of 15 events in the present sample, 11 have $|\Delta t_{i}| < \overline{\Delta t}(15) = 132  $ days. For these 11 interglitch intervals, the mean $\overline{\Delta t}(11) = 14.5  $ days and $\sigma(11)= 131 $ days only, so the model prediction is quite successful. The four events that have $|\Delta t_{i}| > \overline{\Delta t}(15) = 132  $ days all have, suggestively, $ t_{g}$ significantly longer than $t_{obs} $; $344$ days $<$ $\Delta t_{i}$ $<$ $712$ days, with mean $\overline{\Delta t}(4) = 497  $ days and $\sigma(4)= 197 $ days. These 4 glitches will be addressed with an extension of the model, positing that these glitches incorporated ``persistent shifts" in spindown rate as commonly observed in the Crab pulsar's glitches. If there is a persistent shift which is not taken into account, the estimated $t_g$ would be systematically longer than the observed intervals $t_{obs}$, as detailed below, suggesting that the persistent shift may provide the explanation for why all four instances of $> 1 \sigma$ deviations have $t_{g}> t_{obs}$. Without a systematic reason, the random occurrence of $t_{g}> t_{obs}$ in these 4 instances has a probability of 1/16. The idea that this behaviour common in the Crab pulsar glitches can also occur in  some occasional Vela glitches is consistent with the evolution of glitch behaviour with age as one progresses from the Crab pulsar to the Vela pulsar and older pulsars. In the next Section we shall describe and apply a new extended model for predicting the time to the next glitch. 
\begin{table*}
\centering
\caption{The inferred and observed parameters for the Vela glitches. The entries for the first eight glitches and the ninth glitch are taken from \citet{alpar93} and \citet{chauetal93} respectively. Errors are given in parenthesis.The last two columns give the interglitch constant $ \ddot{\Omega} $ and the anomalous braking index $ n_{ig} $.}
   \begin{tabular}{ccccccccccc}
    Year & $ t_{obs} $ (days) & $ t_{g} $ (days) & $\Delta t_{i}$ (days) & $ (I_{A}/I)_{-3} $ & $ (I_{B}/I)_{-3} $ & $ (I_{1}/I)_{-3} $ & $ (I_{2}/I)_{-3} $ & $ (I_{3}/I)_{-3} $ & $ (\ddot{\Omega})_{-21} (rad s^{-3}) $ & $n_{ig}$\\
    \hline
    1969    & 912  & 1624    & 712 & 7.1  &  8.4 & 0.59 & 1.5 & 5.8 & 5.0 & 36.4\\
    1971    & 1491  & 1375    & -116 & 7.2  & 8.8 & 0.59 & 1.5 & 6.4 & 5.9 & 43.6\\
    1975    & 1009  & 1036    & 27 & 7.2  & 12.4 & 0.59 & 1.5 & 5.1 & 7.9 & 57.8\\
    1978    & 1227  & 1371    & 144 &6.6  & 15.2 & 0.59 & 1.5 & 10.0 & 5.5 & 40.0 \\
    1981    & 272  & 616    & 344 & 6.3  & 12.2 & 0.59 & 1.5 & 3.2 & 1.16 & 85.0\\
    1982    & 1067  & 1485    & 418 & 6.0  & 8.4 & 0.59 & 1.5 & 12.1 & 4.6 & 33.6\\
    1985    & 1261  & 972    & -289 & 6.5  & 7.9 & 0.59 & 1.5 & 9.0 & 7.6 & 55.6 \\
    1988    & 907  & 1422    & 515 & 4.7  & 8.2 & 0.59 & 1.5 & 9.5 & 3.7 & 27.5\\
    1991    & 1102  & 1151    & 49 & 7.4  & 16.2 & 0.59 & 1.5 & 10.7 & 7.3 & 53.5\\
    1994    & 778  & 765  & -13 & 5.9(0.03) & 18.8 (0.07) & 0.59 & 1.5 & 8.3 & 8.7 (0.22)& 64.1\\
    1996    & 1190  & 1072    & -118 & 6.6 (0.02)  & 13.1 (0.14) & 0.59 & 1.5 & 12.4 & 7.0 (0.33)& 51.2\\
    2000    & 1634  & 1644     & 10 & 6.4 (0.02) & 12.4 (0.03)& 0.59 & 1.5 & 4.8 & 4.4 (0.21)& 32.4 \\
    2004    & 767  & 913  & 146 & 6.7 (0.02) & 15.8 (0.08) & 0.59 & 1.5  & 7.6 & 8.3 (0.19)& 61.0\\
    2006    & 1449  & 1464   & 15 & 5.0 (0.01) & 12.4 (0.11)& 0.59 & 1.5 & 11.8 & 3.9 (0.23)& 23.4\\
    2010    & 1147  & 1281  & 134 & 6.1 (0.02)  & 25.2 (0.10)& 0.59 & 1.5 & 9.2 & 5.4 (0.17)& 39.6\\
    2013    &  -  &  1553  & - & 6.4 (0.01)  & 13.4 (0.07)& 0.59 & 1.5 & 6.9 & 4.7 (0.09)& 34.2\\
    \hline
    \end{tabular}
\end{table*}

\begin{figure*}
\centering

\subfloat[1994]{\includegraphics[width = 3in,height=4.5cm]{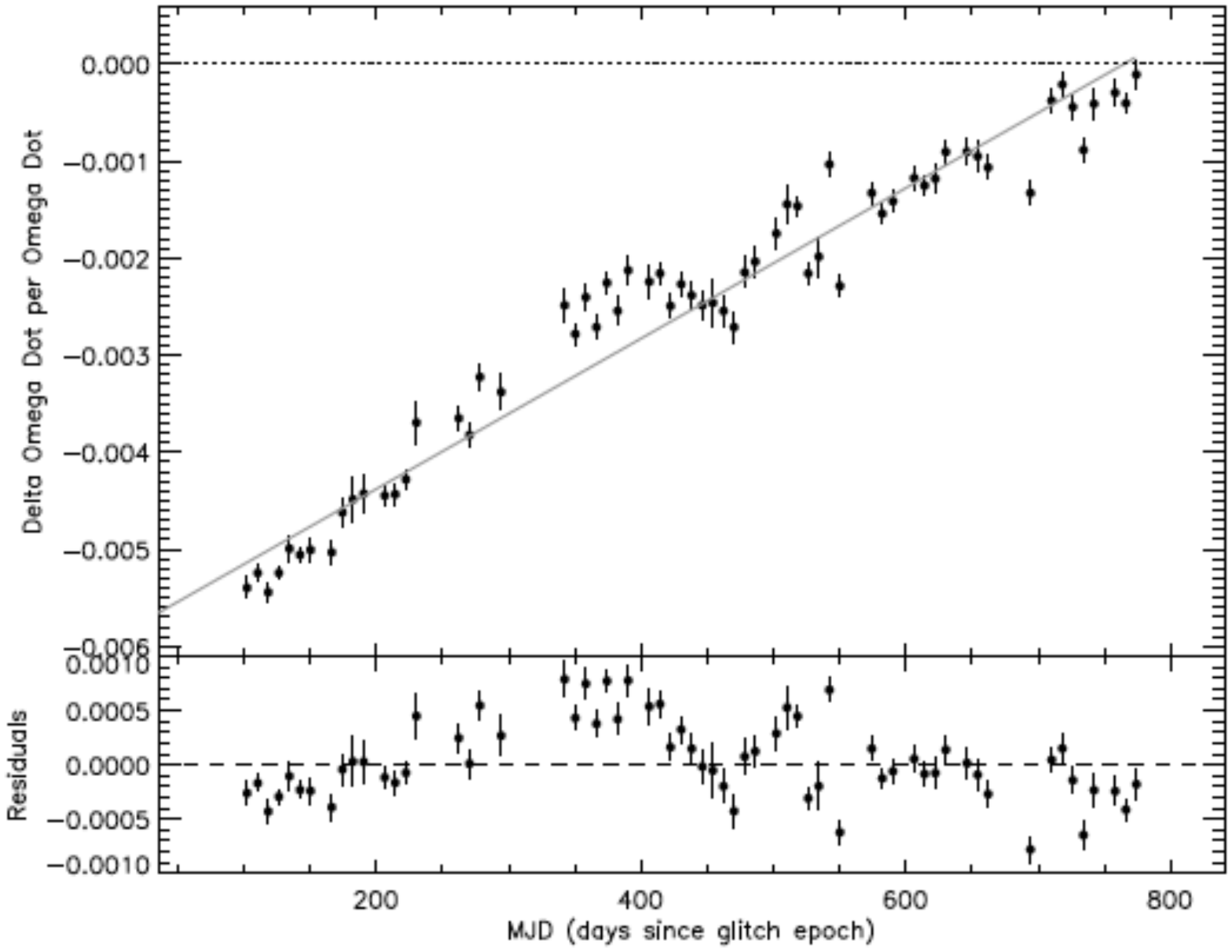}}
\subfloat[1996]{\includegraphics[width = 3in,height=4.5cm]{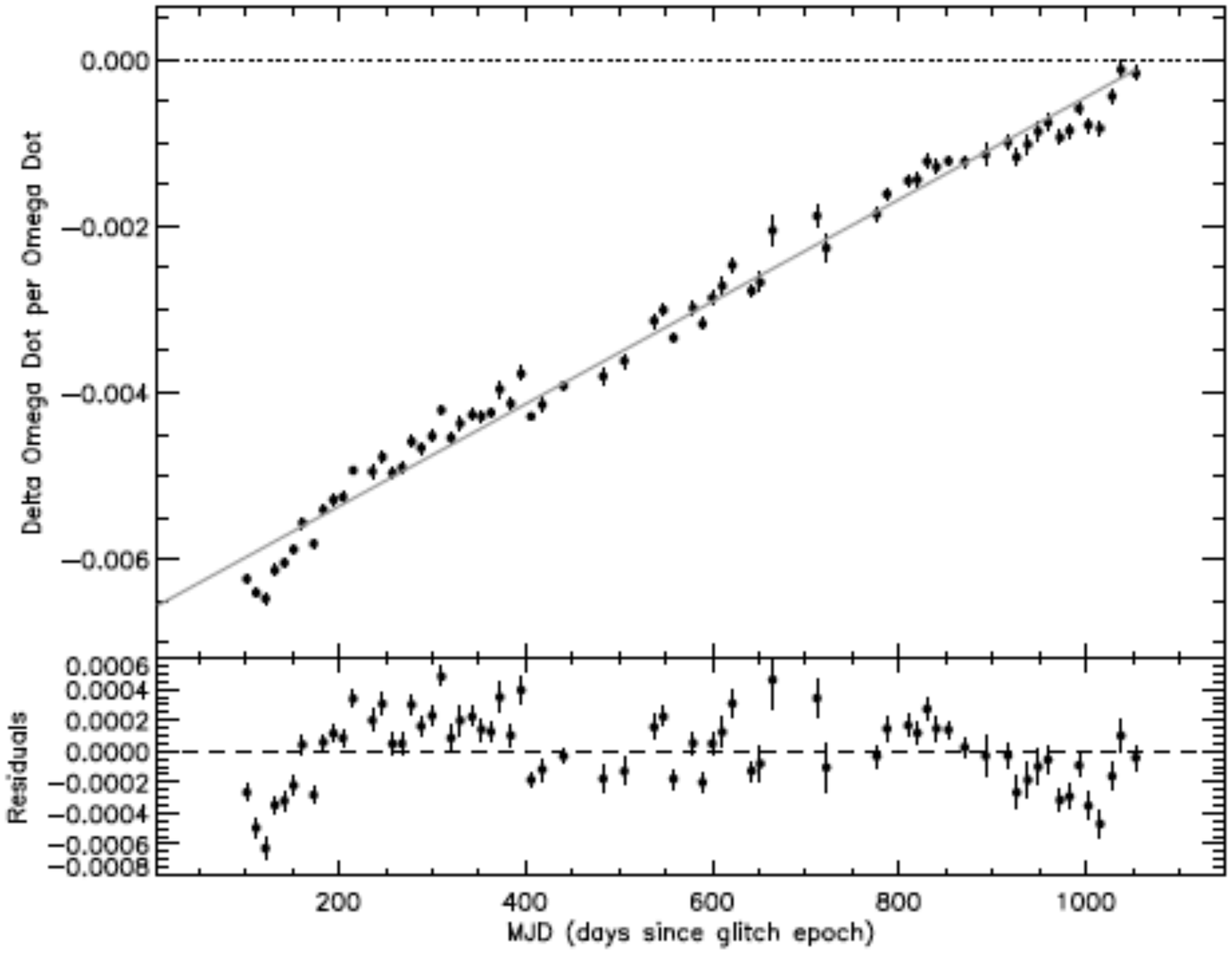}}\
\subfloat[2000]{\includegraphics[width = 3in,height=4.5cm]{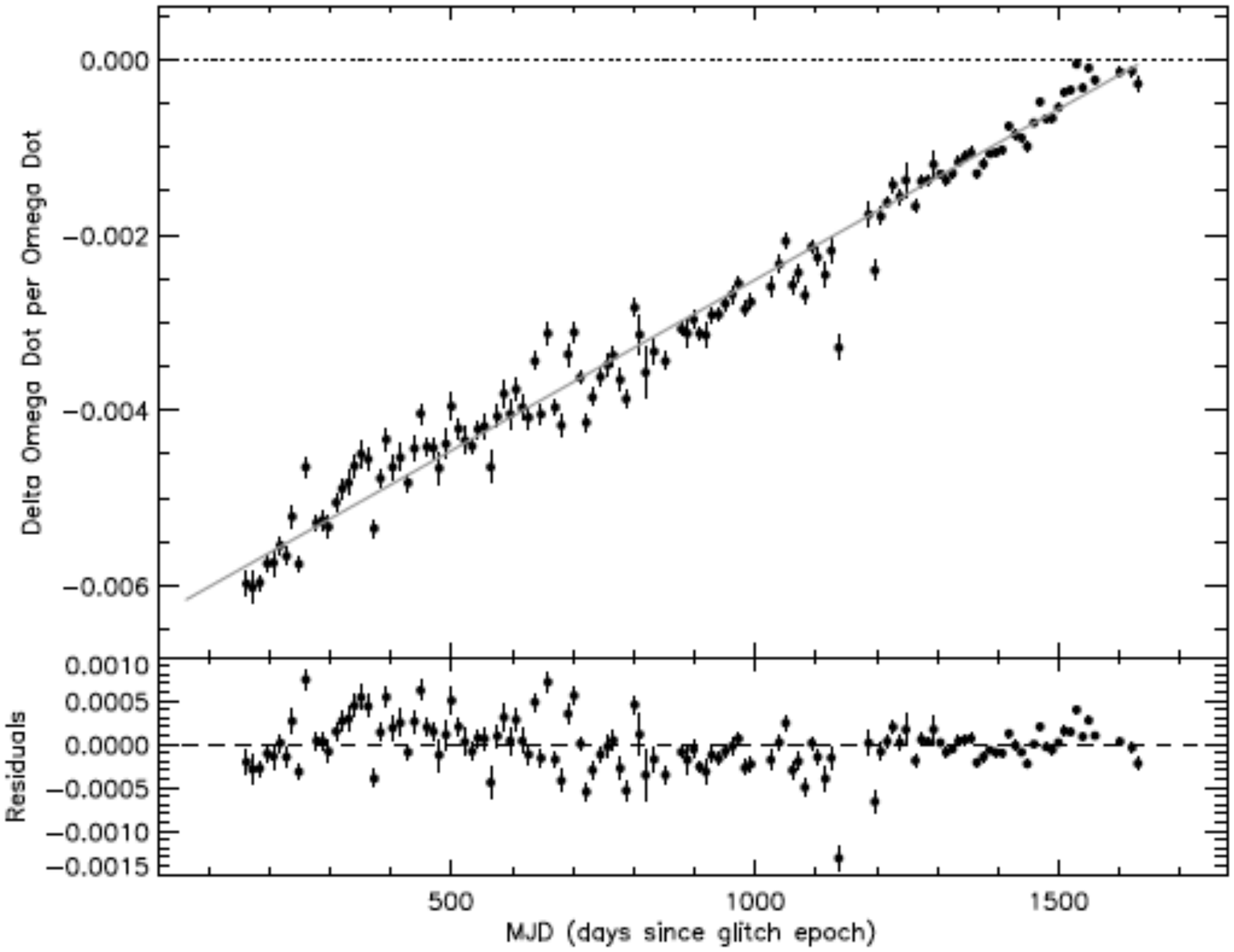}}
\subfloat[2004]{\includegraphics[width = 3in,height=4.5cm]{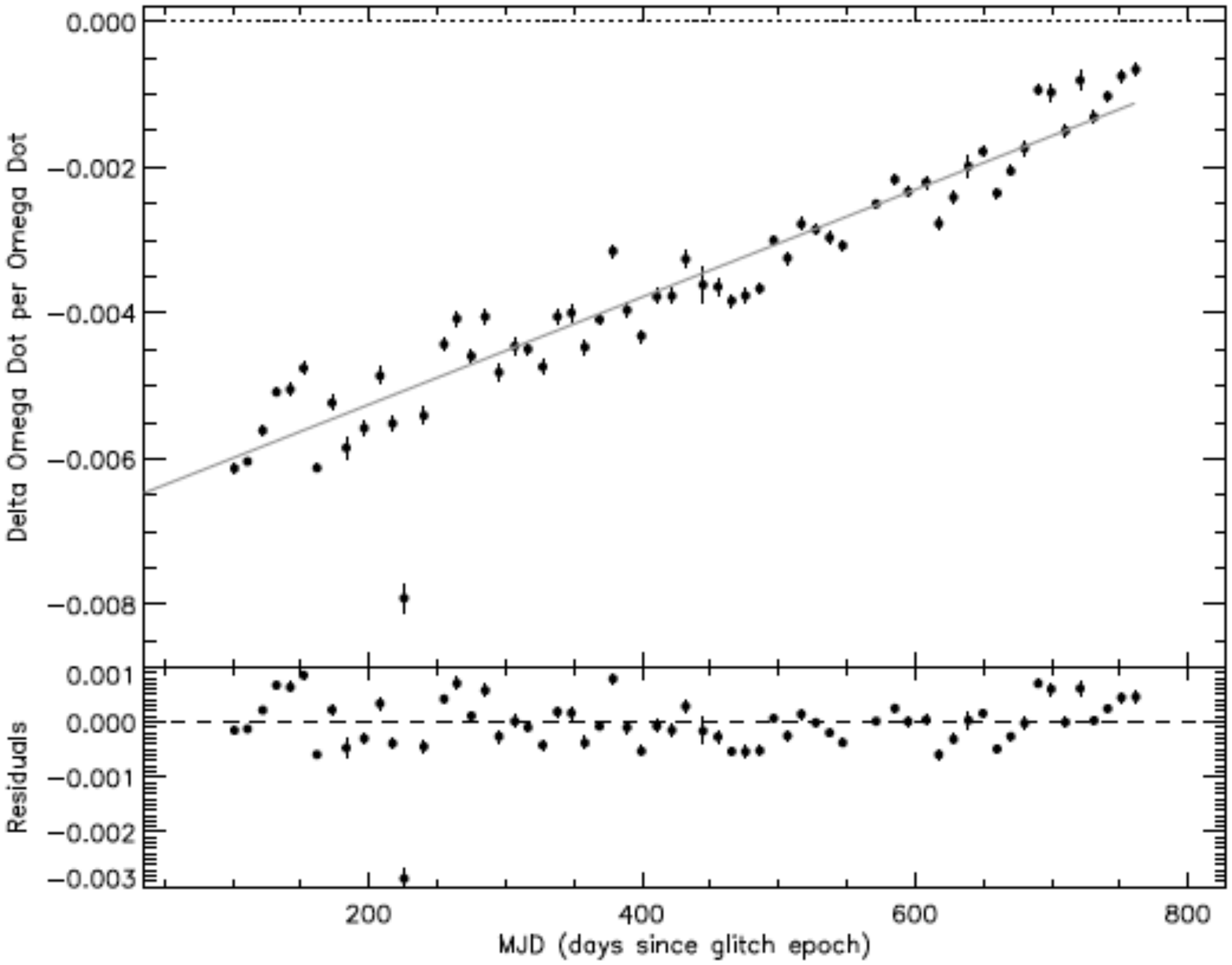}}\
\subfloat[2006]{\includegraphics[width = 3in,height=4.5cm]{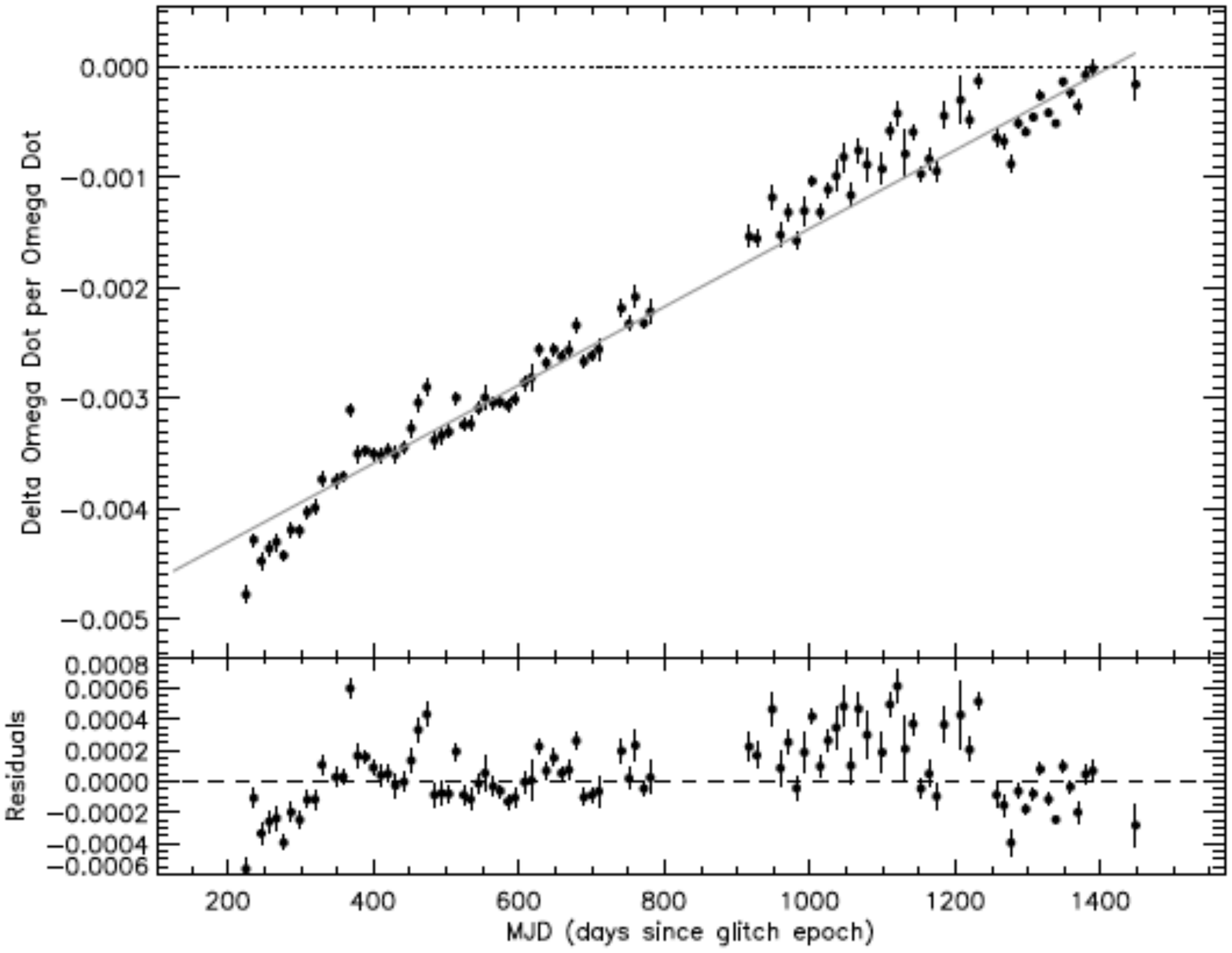}}
\subfloat[2010]{\includegraphics[width = 3in,height=4.5cm]{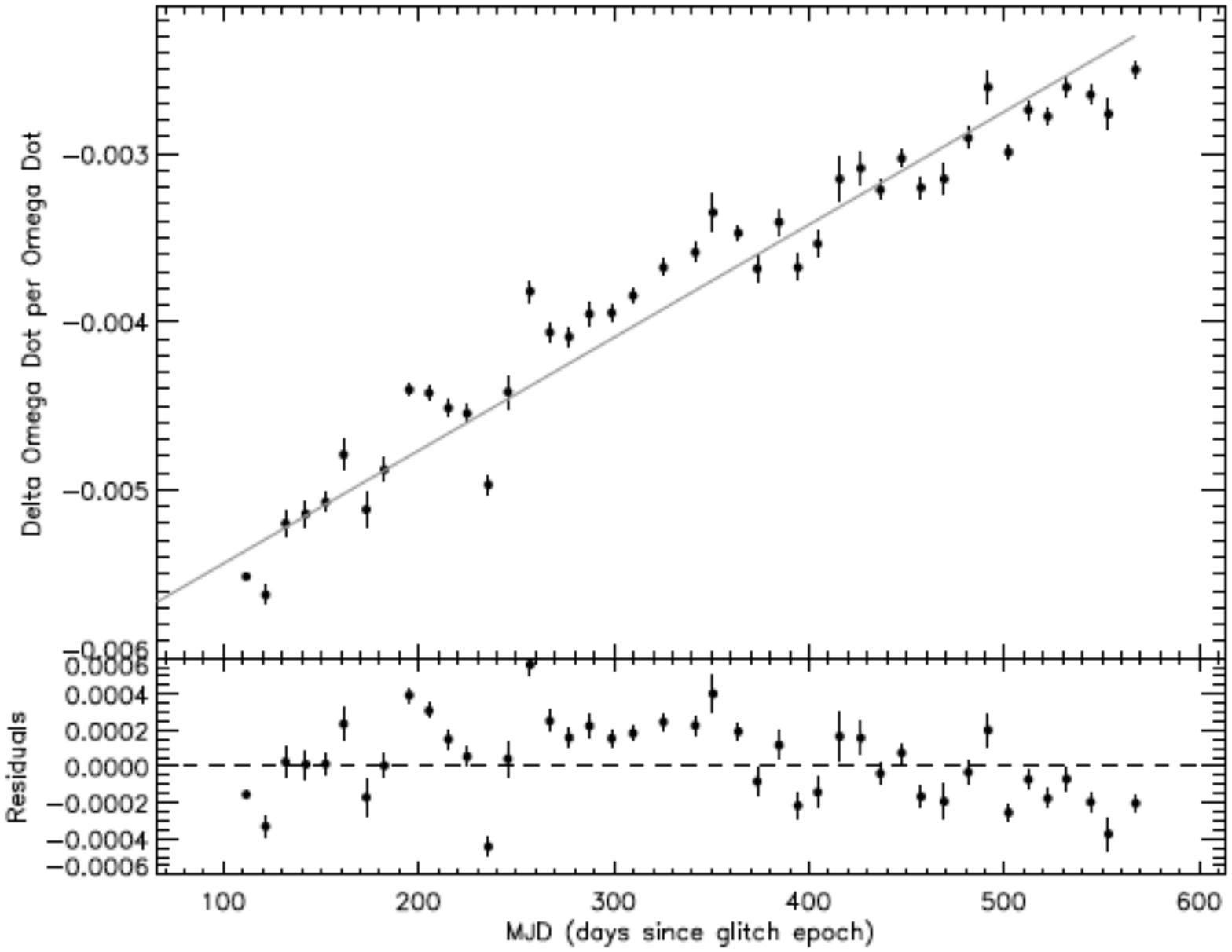}}\
\subfloat[2013]{\includegraphics[width = 3in,height=4.5cm]{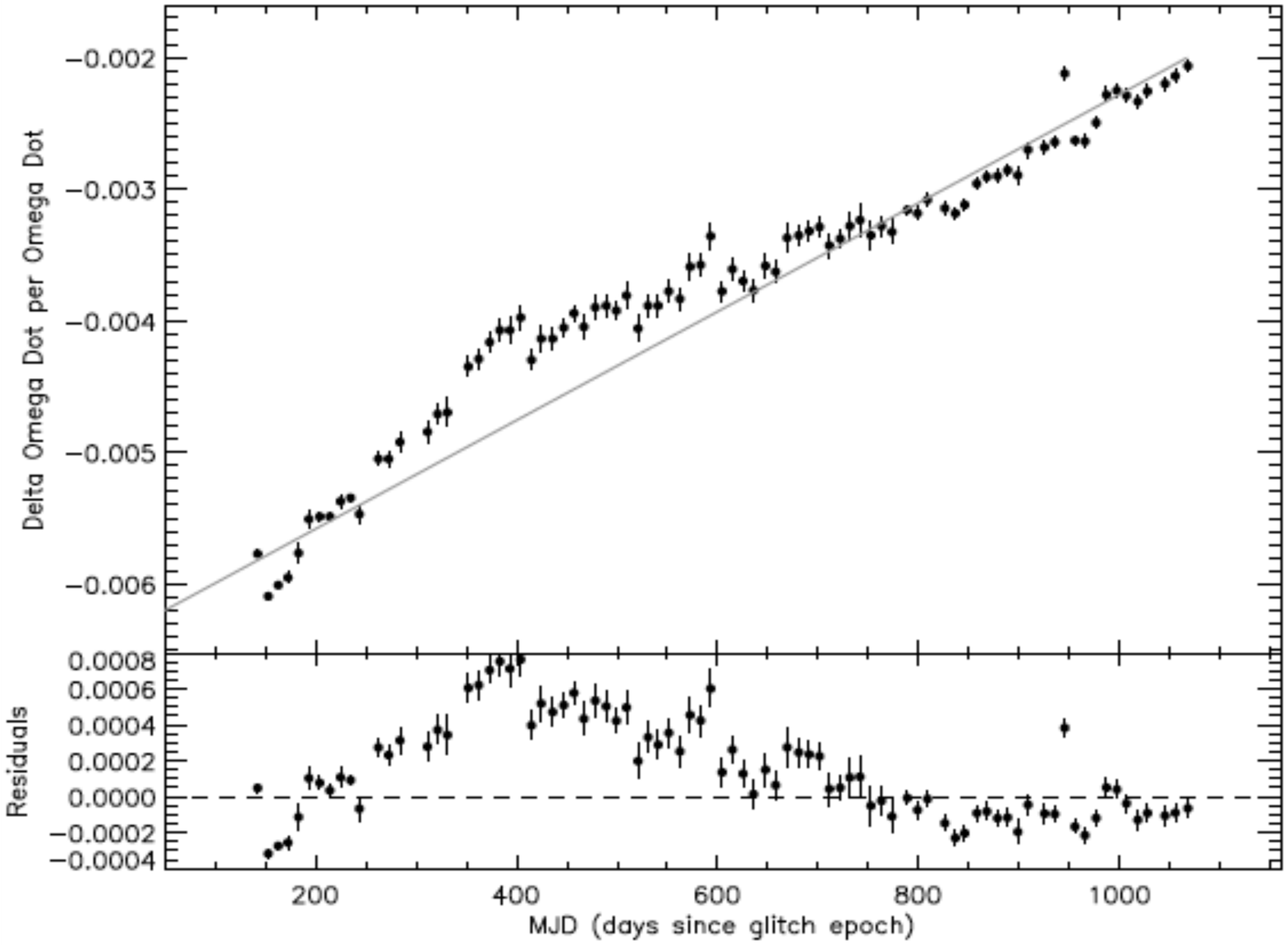}}

\caption{Model fits to post-glitch spindown rate after the 1994, 1996, 2000, 2004, 2006, 2010, 2013 glitches. For the 1994 double glitch, the model fit is applied to data after the second event. The bottom panels show the residuals.}
\label{some example}
\end{figure*}

\section{The Modified Interglitch Times of the Vela Pulsar}\label{sec:interglitch}
The fact that the 4 glitches whose observed arrival times $ t_{obs} $ show substantional discrepancies arrive earlier than the estimated times suggest that recovery is complete when the spindown rate returns to a value somewhat less than its preglitch value. This suggests that each Vela pulsar glitch may be accompanied by a ``persistent shift'', a part of the step in the spindown rate which never recovers, as observed in the Crab pulsar's glitches \citep{lyne15}. For the Crab pulsar, assuming that the glitches are pure unpinning events with creep response leads to estimated $ t_{g} $ of the order of a few months in disagreement with the observed glitch times \citep{alpar96}. This is interpreted as evidence that the glitches in the Crab pulsar are not due to vortex unpinning alone. \cite{alpar96} proposed that the comparatively small ($ \Delta \Omega /\Omega \sim 10^{-8} $) and infrequent ($\sim 6$ yr interglitch time intervals) events in Crab are starquakes \citep{baym71}, in conjunction with unpinning events. In the framework of the vortex creep model, crust cracking is associated with stresses induced by pinned vortices. This scenario also explains the permanent offsets in the spin-down rate,  the ``persistent shifts'', observed in Crab pulsar glitches. The persistent change in spin-down rate is due to newly created vortex traps with surrounding vortex free regions introduced by crust cracking. The newly created vortex free regions were sustaining vortex creep and contributing to spin down before the glitch, while they no longer contribute to $\dot{\Omega}$ after it. The resulting permanent shift in the spin-down rate is given by
\begin{equation}
\frac{\Delta \dot{\Omega}_{p}}{\dot{\Omega}}=\frac{I_{b}}{I}
\end{equation}
where $ \Delta \dot{\Omega}_{p} $ is the observed permanent change in $ \dot{\Omega} $ and $ I_{b} $ is the moment of inertia of the newly created capacitor region. Crust cracking irreversibly restructures the vortex pinning distribution by introducing new capacitor elements. The steady state $ \dot{\Omega} $ value for subsequent glitches is permanently reset, to a new value that is less by $ \Delta \dot{\Omega}_{p} $. The next glitch would then occur roughly when $ \dot{\Omega}$ has returned to $ \Delta \dot{\Omega}_{p} $ less than its steady state value $ \dot{\Omega}_{n,-} $ before the previous glitch: 
\begin{equation}
\dot{\Omega}_{n+1,-}=\dot{\Omega}_{n,-}-\Delta \dot{\Omega}_{p}.
\end{equation}
If such persistent shifts also occur in the Vela pulsar, they would be unresolved in the observed total $ \Delta \dot{\Omega} $ in a glitch. However, the post-glitch non-linear recovery at constant $ \ddot{\Omega} $ would be completed earlier, as shown in Figure 3. We now introduce a modified interglitch time estimator $ t_{g}' $, with the hypothesis that some Vela glitches have a small ``persistent shift'' as observed in the Crab pulsar. 

In the standard vortex creep model \citep{alpar96} in young pulsars like Crab there is a building phase in which vortex traps and the vortex free regions surrounding them are being formed in each glitch, while this building phase is mostly over in the Vela pulsar. In the Crab and Vela pulsars glitches approximately the same number of vortices participate in the glitches. This number, the typical number of vortices unpinning collectively, then acts rather in the manner of a relay race. In the azimuthal direction the unpinned vortices move rapidly relative to the crust lattice through vortex free regions that surround the distribution of vortex traps, while scattering some small distance in the radially outward direction. By the time one batch of unpinned vortices have equilibrated with the superfluid flow they will have ``passed the baton'' by causing the unpinning of a similar number of vortices in vortex traps that were close to critical conditions for unpinning. In the Crab (and other young pulsars) the vortex traps are yet sparse. A batch of unpinned vortices will effect a few other traps on their path, and trigger these traps to unpin, but the avalanche will not travel far, as it fails to find  more traps to unpin. So the total radial distance (or moment of inertia) that the relays of unpinned vortices travel through is only a small fraction of the moment of inertia in the crust superfluid. The size of the glitch depends on the total angular momentum transfer from the superfluid to the normal matter crust. Hence the Crab glitches are much smaller than those in the Vela pulsar. The fact that older pulsars all exhibit glitches that are comparable in size to those of the Vela pulsar suggests that these glitches involve the entire crust superfluid, which represents a connected network for the unpinned vortices to percolate through in relays. The Crab pulsar glitches only involve a small segment of the crust lattice connected to the current site of vortex unpinning, so that the Crab pulsar glitches are not much larger than the minimum glitch size recently resolved \citep{espinoza14}. Thus we arrive at an evolutionary picture, with vortex traps formed in each successive glitch building up from disjoint segment networks of neighbouring vortex traps where unpinning events can travel, as in the Crab pulsar, into a connected network allowing unpinning events to percolate through the entire crust superfluid, defining the maximum glitch size, as observed in Vela and older pulsars. This is supported by the fact that Vela and older pulsars have occasional small Crab-sized glitches while all of the Crab pulsar glitches observed so far have magnitudes  $\Delta\Omega /\Omega \leq 10^{-8}$. Having exhibited two distinct small, Crab sized glitches, the Vela pulsar might also have Crab-like persistent shifts in spin-down rate accompanying some of its glitches. Thus, we are now allowing for the possibility that the Vela pulsar may still be rearranging its network of vortex traps by forming a new vortex trap and surrounding vortex free region during some glitches. 

The part of the $ \Delta \dot{\Omega} $ associated with the permanent shift cannot be discerned at the time of the glitch. Being a persistent shift, it will not relax back. The part of the step down in $ \dot{\Omega} $ that restores with a constant $ \ddot{\Omega} $ will continue the recovery until new steady state conditions that include the persistent shift are reached. A schematic view of the long term behaviour of $ \dot\Omega(t) $ is depicted in Figure 3. In models with persistent shifts, the triangle has less depth, and the estimated time of the next glitch is a bit shorter. This will make the predicted glitch times closer to the observed values. The new estimated glitch time intervals become
\begin{equation}
t_{g}'=t_{g}-\frac{|\Delta \dot{\Omega}_{p}|}{\ddot{\Omega}} = \frac{|\Delta \dot{\Omega}|}{\ddot{\Omega}} - \frac{|\Delta \dot{\Omega}_{p}|}{\ddot{\Omega}}
\end{equation}
where $ \Delta \dot{\Omega} $ and $ \Delta \dot{\Omega}_{p} $ is the total and permanent step in spindown rate respectively.

\begin{figure}
\centering
\vspace{0.1cm}
\includegraphics[width=1.0\linewidth]{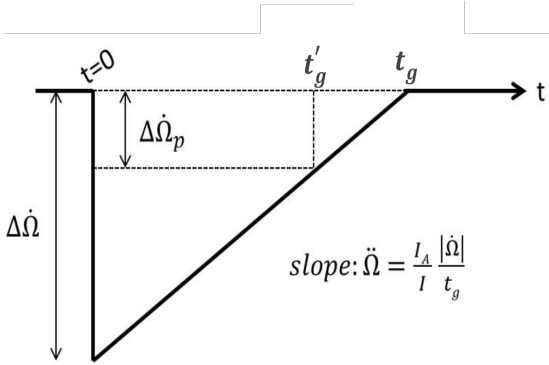}
\caption{The schematic view of the long term behaviour of $ \Delta \dot{\Omega} $.}
\label{fig:omegadot}
\end{figure}

The values of $ \Delta \dot{\Omega}_{p} / \dot{\Omega} $ are between $ (0.2-6.6) \times 10^{-4} $ for Crab pulsar glitches \citep{lyne15}. It may be different in Vela due to evolutionary reasons. We focus on the four glitches, for which the estimated time $t_{g}$ to the next glitch was considerably longer than the observed interval $t_{obs}$.  As Model (2), we assume that each of these four glitches had a persistent shift of the same fractional magnitude as the average persistent shift for the Crab pulsar glitches.
\begin{equation}
\frac{\Delta \dot{\Omega}_{p}}{\dot{\Omega}} = \langle \frac{\Delta \dot{\Omega}_{p}}{\dot{\Omega}}\rangle_{Crab} = 2.5 \times 10^{-4} \nonumber
\end{equation}

The new estimates $ t_{g}' $ and $\Delta t_i$ are given in Table 2. The mean and standard deviation for the 4 glitches are now $\overline{\Delta t}(4) = 443  $ days and $\sigma(4)= 150 $ days, while for the full sample of 15 glitches $\overline{\Delta t}(15) = 117  $ days and the standard deviation $\sigma(15)= 241 $ days. Introducing persistent shifts of the same magnitude as observed in the Crab pulsar is clearly not enough to make better estimates of the arrival times of the next glitch. We next consider the possibility that these five Vela glitches involved a major restructuring of the vortex trap network so that $t_g$ was reduced by a third. The estimates $ t_{g}' = (1 - 0.33) t_{g} $, Model (3), given in Table 3, lead to $\overline{\Delta t}(4) = 73 $ days and $\sigma(4)= 152 $ days, while for the full sample of 15 glitches $\overline{\Delta t}(15) = 18  $ days and the standard deviation $\sigma(15)= 127 $ days. Tables 2 and 3 also give the model parameters $ I_{A}/I $ and $ I_{B}/I $. 
We have applied Kolmogorov-Smirnov (KS) tests to Model(1), without persistent shifts, and Model(3), with persistent shifts corresponding to a substantial rearrangement of the vortex creep regions. We first construct a cumulative distribution for the observed interglitch intervals $t_{obs}$, without binning. Comparing the cumulative distribution of our Model(1) estimates $t_{g}$ with the $t_{obs}$ cumulative distribution, we find that the null hypothesis is rejected at a confidence level of only 39 \% . Comparing the cumulative distribution of the estimates $t_{g}'$ of Model (3) with the cumulative $t_{obs}$ distribution, a  confidence level of 98 \% is obtained. We conclude that our Model(3) with large persistent shifts produces substantial agreement with the observed glitch time intervals. The simulations of \citet{MelatosWarszawski09} and \citet{WarszawskiMelatos11} also suggest that a substantial fraction of vortex traps and vortex free regions are involved and rebuilt in each glitch.

\begin{table}
\centering
\caption{Estimations of the model parameters for the four glitches for which $t_{g}$ is considerably longer than $t_{obs}$ if the persistent shift is the same magnitude as the average value for Crab, $ \Delta \dot{\Omega}_{per}/\dot{\Omega} = 2.5 \times 10^{-4}$.}
\medskip\medskip
   \begin{tabular}{cccccc}
    Year & $ t_{obs}$ (days) & $ t_{g}' $ (days)  & $\Delta t_{i}$ (days) & $ (I_{A}/I)_{-3} $ & $ (I_{B}/I)_{-3} $ \\
    \hline
    1969    & 912  & 1567     & 655  &  6.9 & 9.0 \\
    1981    & 272  & 592    & 320  & 6.1 & 12.9 \\
    1982    & 1067  & 1423    & 356  & 5.8 & 9.1 \\
    1988    & 907  & 1346   & 439  & 4.5 & 8.9 \\
    \hline
    \end{tabular}
\end{table}

\begin{table*}
\centering
\caption{Estimations of the model parameters for the four glitches for which $t_{g}$ is considerably longer than $t_{obs}$ if $ t_{g}' = (1 - 0.33) t_{g} $ (Model 3).}
\medskip\medskip
   \begin{tabular}{ccccccc}
    Year & $ t_{obs} $ (days) & $ t_{g}' $ (days) & $\Delta t_{i}$ (days) &  $ (\Delta \dot{\Omega}_{per}/\dot{\Omega})_{-3} $  & $ (I_{A}/I)_{-3} $ & $ (I_{B}/I)_{-3} $ \\
    \hline
    1969    & 912  & 1088    &  176 & 2.3  &  4.8 & 15.5 \\
    1981    & 272  & 413    &  141 & 2.4  & 3.9 & 21.1 \\
    1982    & 1067  & 995    & -72 & 2.3  & 3.7 & 15.3 \\
    1988    & 907  & 953    & 46 & 2.3  & 2.4 & 14.5\\
    \hline
    \end{tabular}
\end{table*}

\section{The Braking Index of the Vela Pulsar} \label{sec:brakingindex}

We now use the above results to calculate the braking index for the Vela pulsar. Estimating the braking index requires modeling and subtracting all contributions of glitches and post-glitch and interglitch recovery. \cite{lyne96} did this by assuming that $ \dot{\Omega} $ values 150 days after each glitch are already clean of post-glitch response and estimated a braking index $ n\cong 1.4 $. They used a post-glitch epoch when all short term exponential relaxation components with $ \tau_{1}=10 $ hr, $ \tau_{2}=3.2 $ days and $ \tau_{3}=32 $ days are over, but the inter-glitch recovery is far from finished. However, one also needs to  take account of the constant $ \ddot{\Omega} $ response, which extends to the next glitch. The appropriate epochs when inter-glitch response is completed should be those that are  just prior to the subsequent glitch. \cite{espinoza16} take into account the constant $ \ddot{\Omega} $ term, but assume that this recovery is not completed before the next glitch. The data fits in this work do not extend to the end of the interglitch interval. Instead fiducial epochs are chosen by extrapolating the $ \dot{\nu}(t) $ ($ \dot{\Omega} $) fits back to immediately after the previous glitch. They obtain $ n \cong 1.7 \pm 0.2 $, and suggest that this low braking index is due to incomplete recovery of the glitch in the spin-down rate. But the recovery continues up to, and is actually completed at, a time just before the next glitch.   

In those glitches for which timing observations immediately before the next glitch are not available we obtain the $ \dot{\Omega}(t_{obs,i}) $ values at the epochs immediately prior to the next glitch by extrapolating from our fits with constant $ \ddot{\Omega} $. We then produce the very long term (between the years 1969-2013) $ \ddot{\Omega}_{PSR} $ values for the Vela pulsar by the best linear fit to the $ \dot{\Omega}(t_{obs,i}) $. This gives the estimate of  $ \ddot{\Omega}_{PSR}^{t_{obs}} = (3.83\pm 0.15) \times 10^{-22} $ rad s$ ^{-3} $ leading to the braking PSR index $ n_{t_{obs}}= 2.81 \pm 0.12 $ for the Vela pulsar, with
\begin{equation}
n=\frac{\overline{\Omega} \ddot{\Omega}_{PSR}}{\left( \overline{\dot{\Omega}} \right)^{2}},
\end{equation} 
where $\overline{\Omega}=70.4 $ rad s$^{-1}$ and $\overline{\dot{\Omega}}=9.8 \times 10^{-11}$ rad s$ ^{-2} $ are the average values over the $ 47 $ year data span. The quoted errors in the braking index are propagated from the errors in our long term fit for $ \ddot{\Omega}_{PSR} $. The braking index 
$n_{t_{obs}}= 2.81 \pm 0.12 $, using spindown rates just prior to the observed time of the next glitch, is independent of any estimations of the glitch time. It is based only on the premises that the approximately constant second derivative of the rotation rate observed at interglitch epochs (i) is due to internal torques, and (ii) determines the value of the spin- down rate just prior to the next glitch. This is supported by all the instances where timing data extend up to the next glitch. These data sets are well described by a constant $\ddot{\Omega}$ all the way to the next glitch, without any precursor signatures prior to the glitch. This justifies using extrapolations with the constant $\ddot{\Omega}$ in those inter-glitch data sets that do not extend to the next glitch.

Our other estimates of the braking index pick an estimated time of the next glitch (16 estimates including the next glitch after 2013) and are thus model dependent. Model (3) which gives a better estimations of glitch times leads to $ n_{t_{g}'}=2.87 \pm 0.17 $. Within error bars, this is consistent with $ n_{t_{obs}}= 2.81 \pm 0.12 $.

\begin{figure}
\centering
\includegraphics[width=1.15\linewidth]{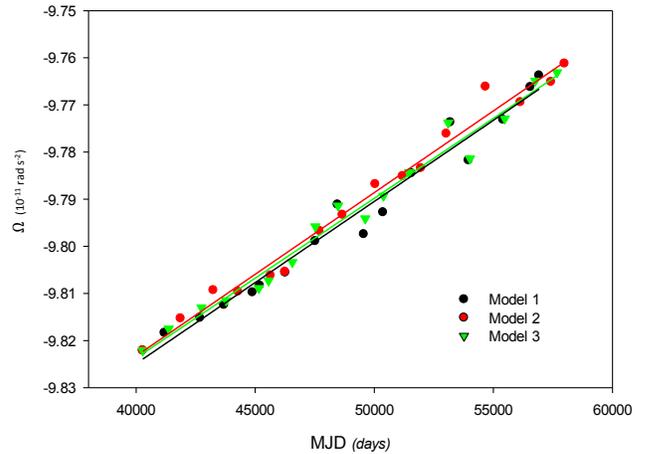}
\caption{Spin-down rates at the fiducial epochs $ t_{obs} $ (black points, Model 1), $ t_{g}' $ (red points, Model 2), and $ t_{g}' $ (green points, Model 3) prior to the next glitch (1971-2013). The best straight line fits used to extract the braking index $ n $ are also shown.}
\label{fig:brakingindex}
\end{figure}

\section{Conclusions }
\label{sec:discussion}
We have confirmed that the interglitch timing behaviour of the Vela pulsar in the cumulative data from discovery to 2016, covering 16 glitches is characterized by a recovery of the spin-down rate at a constant $\ddot{\Omega}$, as was seen in the earlier data up to the 9th glitch. Power law behaviour, like the constant $\ddot{\Omega}$, is a clear signal of nonlinear dynamics. This behaviour results from the cumulative manifestation of the step like (``Fermi function'') recovery of the spindown rate predicted by the vortex creep model \citep{alpar84a} that has been directly observed in the Vela pulsar \citep{flanagan95, buchner08}. It is expected to determine the inter-glitch timing once the linear response in the form of exponential relaxations is over. The interglitch second derivative $\ddot{\Omega}$ of the rotation rate is determined by the parameters of the previous glitch. The completion of this ubiquitous interglitch behaviour signals the arrival of the next glitch and allows for an estimate of the time of the next glitch. We have shown that the constant Ω ̈ behaviour dominates the interglitch interval for the full Vela pulsar data set comprising 16 glitches so far. Beyond the phenomenology, for 11 of the 15 intervals between glitches $\ddot{\Omega}$ relates to the glitch parameters in the way predicted by vortex creep theory, leading to predictions of inter-glitch intervals that agree with the observed glitch dates within 12 \%. This level of agreement between theory and observation represents  strong evidence in support of the non-linear vortex creep theory. 

The rms residuals from the constant $ \ddot{\Omega} $ fits are $ \lesssim $ 5 \% of $ \Delta \dot{\Omega} $. These deviations correspond to spatial fluctuations from a uniform density of vortices unpinned in the glitch which gives rise to the constant $ \ddot{\Omega} $ recovery.   

With the daily monitoring of the Vela pulsar at Hartebesthoek Observatory (HartRAO) timing data extending to the next glitch are available for inter-glitch intervals since 1985, confirming the constant $\ddot{\Omega}$ behaviour. In those cases where there is a gap in the data prior to the next glitch, we have extrapolated the inter-glitch data to the epoch prior to the next glitch. Using observed or extrapolated values of the spindown rate at the epochs prior to the glitches, we arrived at a braking index $ n = 2.81 \pm 0.12 $. This value is in agreement with the braking indices $n \leq 3 $ observed for most isolated pulsars \citep{melatos97, lyne15, antonopoulou15, archibald15, clark16}.

Furthermore, we find there is an explanation of the 4 cases for which the inter-glitch intervals predicted by the standard vortex creep theory are considerably longer than the observed intervals, with a mean fractional offset $\overline{\Delta t}(4) = 497 $ days and standard deviation $\sigma(4)= 197 $ days. We propose that for each of these glitches there might be a persistent shift in spin-down rate that does not relax back entirely. This step which is commonly observed in the Crab pulsar glitches, would not be distinguished observationally as a part of the glitch in the spin-down rate. We introduce the possibility that some Vela glitches may involve the making of new vortex traps and vortex free regions, as postulated earlier to explain the Crab pulsar's persistent shifts within vortex creep theory. After each glitch these newly restructured regions no longer contribute to spin-down and cause a permanent shift in spin-down rate. The triangle inter-glitch behaviour in spin-down, which restores with a constant $\ddot{\Omega}$, is then completed somewhat earlier, restoring conditions for a new glitch. Deriving the modified estimator for the time between glitches by using this consideration, with persistent shifts of the average magnitude observed in the Crab pulsar, our Model (2), does not result in appreciable improvement of the estimates of the interglitch intervals. If we assume, as in Model (3), that as much as a third of the Vela pulsar's vortex creep network is restructured in each of these four glitches, the new interglitch time estimates after the four glitches have a mean fractional offset of only $\overline{\Delta t}(4) = 73 $ days and standard deviation $\sigma(4)= 152 $ days for the full sample of 15 glitches with an observed interval to the next glitch. It is suggestive that for all four glitches with $t_g$ significantly longer than $t_{obs}$ the idea that some glitches involve restructuring of the vortex creep network leads to better estimation. We do not have an understanding of why these particular 4 glitches involve rather substantial rearrangements, but this may be only natural in a complex system with nonlinear dynamics where the glitches are instabilities of the pinned vortex distribution, possibly triggered in interaction with lattice stresses. We use Models (2) and (3) as estimates of the range in predicted arrival times of the glitches, leading to a corresponding range of the braking index, $ n = 2.93 \pm 0.15 $ for Model (2) and $ n = 2.87 \pm 0.17 $ for Model (3). These are consistent, within error bars, with the value $ n = 2.81 \pm 0.12 $ obtained by using extrapolations to the epochs just prior to the next glitch. 

We can now present predictions for the time of the glitch according to the vortex creep model. The next glitch will occur $ t_{g} = 1553 \pm 132 $ days after the last observed glitch of 2013, around Dec. 22, 2017, if no persistent shift was associated with the 2013 glitch. If a persistent shift of the order of those observed in the Crab pulsar occurred then the glitch interval is $ t_{g} = 1514 \pm 183 $ days, predicting the arrival of the next glitch around Nov. 11, 2017. If the 2013 glitch induced a large persistent shift associated with major restructuring of the vortex creep network, the next glitch will take place as early as July 27, 2016, within an interval of $ t_{g} = 1040 \pm 152 $ days. The uncertainties for each case correspond to one standard deviation.

An approximately constant second derivative of the rotation rate correlated with glitch parameters is also seen in the postglitch or interglitch timing of most older pulsars which exhibit Vela-like giant glitches \citep{yu13}. Furthermore the ``anomalous'' braking indices observed from many pulsars \citep{johnston99} can be understood as due to nonlinear creep response, with rotation rate second derivatives correlated with glitch parameters similar to those of the Vela pulsar \citep{alpar06}. The long term interglitch timing of the Crab pulsar covering all glitches so far \citep{cades16, lyne15} and the large glitch associated change in the reported braking index of PSR J1846-0258 \citep{archibald15} are also associated with the interglitch $\ddot{\Omega}$ predicted by the nonlinear vortex creep model, as we will discuss in future work. 

Literally while we were submitting the first version of this paper we saw ATel \# 9847 from Jim Palfreyman of the University of Tasmania, announcing that the Vela pulsar had glitched on that day, Dec. 12, 2016 \citep{palfreyman16}. The glitch has arrived 138 days late, within the $1 \sigma$ uncertainty of $152$ days from the prediction of our model with a persistent shift.
   
\section*{Acknowledgments}

This work is supported by the Scientific and Technological Research Council of Turkey (T\"{U}B\.{I}TAK) under the grant 113F354. M.A.A. and D.P. thank the Aspen Center of Physics, which is supported by National Science Foundation grant PHY-1066293, for its hospitality during the period they collaborated on this paper, and M.A.A. thanks the Simons Foundation for a grant that made his visit to the Aspen Center for Physics possible. M.A.A. is a member of the Science Academy (Bilim Akademisi), Turkey. The data presented herein were obtained at the Hartebeesthoek Radio Astronomy Observatory (HartRAO). HartRAO is a National Facility of  the National Research Foundation (NRF) of South Africa. We would like to thank the directors and staff of HartRAO for supporting the Vela monitoring program for the last 32 years, and especially Claire Flanagan  who started the Vela observing program and detected the first evidence for nonlinear response in the form of a step recovery of the spindown rate. We thank Erbil G\"ugercino\u{g}lu for useful discussions and Jim Palfreyman for alerting us to a mistake in the first version of the paper and for communications on the Dec. 12, 2016 glitch.   


\bibliographystyle{mn2e}
\bibliography{vela_ref}

\providecommand{\noopsort}[1]{}\providecommand{\singleletter}[1]{#1}%
\begin{thebibliography}{51}
\expandafter\ifx\csname natexlab\endcsname\relax\def\natexlab#1{#1}\fi

\bibitem[{{Alpar}(1977)}]{alpar77}
{Alpar} M.~A., 1977, apj, 213, 527

\bibitem[{{Alpar} {et~al}\mbox{.}(1984{\natexlab{a}}){Alpar}, {Anderson},
  {Pines}, \& {Shaham}}]{alpar84a}
{Alpar} M.~A., {Anderson} P.~W., {Pines} D., {Shaham} J., 1984{\natexlab{a}},
  apj, 276, 325

\bibitem[{{Alpar} {et~al}\mbox{.}(1984{\natexlab{b}}){Alpar}, {Anderson},
  {Pines}, \& {Shaham}}]{alpar84b}
{Alpar} M.~A., {Anderson} P.~W., {Pines} D., {Shaham} J., 1984{\natexlab{b}},
  apj, 278, 791

\bibitem[{{Alpar} \& {Baykal}(1994)}]{alpar94}
{Alpar} M.~A., {Baykal} A., 1994, mnras, 269, 849

\bibitem[{{Alpar} \& {Baykal}(2006)}]{alpar06}
{Alpar} M.~A., {Baykal} A., 2006, mnras, 372, 489

\bibitem[{{Alpar} {et~al}\mbox{.}(1993){Alpar}, {Chau}, {Cheng}, \&
  {Pines}}]{alpar93}
{Alpar} M.~A., {Chau} H.~F., {Cheng} K.~S., {Pines} D., 1993, apj, 409, 345

\bibitem[{{Alpar} {et~al}\mbox{.}(1996){Alpar}, {Chau}, {Cheng}, \&
  {Pines}}]{alpar96}
{Alpar} M.~A., {Chau} H.~F., {Cheng} K.~S., {Pines} D., 1996, apj, 459, 706

\bibitem[{{Alpar}, {Cheng} \& {Pines}(1989){Alpar}, {Cheng}, \&
  {Pines}}]{alpar89}
{Alpar} M.~A., {Cheng} K.~S., {Pines} D., 1989, apj, 346, 823

\bibitem[{{Anderson} \& {Itoh}(1975)}]{anderson75}
{Anderson} P.~W., {Itoh} N., 1975, nat, 256, 25

\bibitem[{{Antonopoulou} {et~al}\mbox{.}(2015){Antonopoulou}, {Weltevrede},
  {Espinoza}, {Watts}, {Johnston}, {Shannon}, \& {Kerr}}]{antonopoulou15}
{Antonopoulou} D., {Weltevrede} P., {Espinoza} C.~M., {Watts} A.~L., {Johnston}
  S., {Shannon} R.~M., {Kerr} M., 2015, mnras, 447, 3924

\bibitem[{{Archibald} {et~al}\mbox{.}(2015){Archibald}, {Kaspi}, {Beardmore},
  {Gehrels}, \& {Kennea}}]{archibald15}
{Archibald} R.~F., {Kaspi} V.~M., {Beardmore} A.~P., {Gehrels} N., {Kennea}
  J.~A., 2015, apj, 810, 67

\bibitem[{{Baym}, {Pethick} \& {Pines}(1969){Baym}, {Pethick}, \&
  {Pines}}]{baym69}
{Baym} G., {Pethick} C., {Pines} D., 1969, nat, 224, 673

\bibitem[{{Baym} \& {Pines}(1971)}]{baym71}
{Baym} G., {Pines} D., 1971, Annals of Physics, 66, 816

\bibitem[{{Buchner}(2013)}]{buchner13}
{Buchner} S., 2013, The Astronomer's Telegram, 5406

\bibitem[{{Buchner} \& {Flanagan}(2008)}]{buchner08}
{Buchner} S., {Flanagan} C., 2008, in American Institute of Physics Conference
  Series, Vol. 983, 40 Years of Pulsars: Millisecond Pulsars, Magnetars and
  More, {Bassa} C., {Wang} Z., {Cumming} A., {Kaspi} V.~M., eds., pp. 145--147

\bibitem[{{Buchner} \& {Flanagan}(2011)}]{buchner11}
{Buchner} S., {Flanagan} C., 2011, in American Institute of Physics Conference
  Series, Vol. 1357, American Institute of Physics Conference Series, {Burgay}
  M., {D'Amico} N., {Esposito} P., {Pellizzoni} A., {Possenti} A., eds., pp.
  113--116

\bibitem[{{Buchner}(2010)}]{buchner10}
{Buchner} S.~J., 2010, The Astronomer's Telegram, 2768

\bibitem[{{Chau} {et~al}\mbox{.}(1993){Chau}, {McCulloch}, {Nandkumar}, \&
  {Pines}}]{chauetal93}
{Chau} H.~F., {McCulloch} P.~M., {Nandkumar} R., {Pines} D., 1993, apjl, 413,
  L113

\bibitem[{{Clark} {et~al}\mbox{.}(2016){Clark}, {Pletsch}, {Wu}, {Guillemot},
  {Camilo}, {Johnson}, {Kerr}, {Allen}, {Aulbert}, {Beer}, {Bock},
  {Cu{\'e}llar}, {Eggenstein}, {Fehrmann}, {Kramer}, {Machenschalk}, \&
  {Nieder}}]{clark16}
{Clark} C.~J. {et~al.}, 2016, apjl, 832, L15

\bibitem[{{Datta} \& {Alpar}(1993)}]{datta93}
{Datta} B., {Alpar} M.~A., 1993, aap, 275, 210

\bibitem[{{Dodson} {et~al}\mbox{.}(2004){Dodson}, {Buchner}, {Reid}, {Lewis},
  \& {Flanagan}}]{dodson04}
{Dodson} R., {Buchner} S., {Reid} B., {Lewis} D., {Flanagan} C., 2004, iaucirc,
  8370

\bibitem[{{Dodson}, {McCulloch} \& {Lewis}(2002){Dodson}, {McCulloch}, \&
  {Lewis}}]{dodson02}
{Dodson} R.~G., {McCulloch} P.~M., {Lewis} D.~R., 2002, apjl, 564, L85

\bibitem[{{Edwards}, {Hobbs} \& {Manchester}(2006){Edwards}, {Hobbs}, \&
  {Manchester}}]{edwards06}
{Edwards} R.~T., {Hobbs} G.~B., {Manchester} R.~N., 2006, mnras, 372, 1549

\bibitem[{{Espinoza} {et~al}\mbox{.}(2014){Espinoza}, {Antonopoulou},
  {Stappers}, {Watts}, \& {Lyne}}]{espinoza14}
{Espinoza} C.~M., {Antonopoulou} D., {Stappers} B.~W., {Watts} A., {Lyne}
  A.~G., 2014, mnras, 440, 2755

\bibitem[{{Espinoza}, {Lyne} \& {Stappers}(2017){Espinoza}, {Lyne}, \&
  {Stappers}}]{espinoza16}
{Espinoza} C.~M., {Lyne} A.~G., {Stappers} B.~W., 2017, mnras, 466, 147

\bibitem[{{Flanagan}(1991)}]{flanagan91}
{Flanagan} C., 1991, iaucirc, 5311

\bibitem[{{Flanagan} \& {McCulloch}(1994)}]{flanagan94}
{Flanagan} C., {McCulloch} P.~M., 1994, iaucirc, 6038

\bibitem[{{Flanagan}(1995)}]{flanagan95}
{Flanagan} C.~S., 1995, in NATO Advanced Science Institutes (ASI) Series C,
  Vol. 450, NATO Advanced Science Institutes (ASI) Series C, {Alpar} M.~A.,
  {Kiziloglu} U., {van Paradijs} J., eds., p. 181

\bibitem[{{Flanagan} \& {Buchner}(2006)}]{flanagan06}
{Flanagan} C.~S., {Buchner} S.~J., 2006, Central Bureau Electronic Telegrams,
  595

\bibitem[{{G{\"u}gercino{\u g}lu} \& {Alpar}(2014)}]{guger14}
{G{\"u}gercino{\u g}lu} E., {Alpar} M.~A., 2014, apjl, 788, L11

\bibitem[{{Haskell} \& {Antonopoulou}(2014)}]{haskell14}
{Haskell} B., {Antonopoulou} D., 2014, mnras, 438, L16

\bibitem[{{Hobbs}, {Edwards} \& {Manchester}(2006){Hobbs}, {Edwards}, \&
  {Manchester}}]{hobbs06}
{Hobbs} G.~B., {Edwards} R.~T., {Manchester} R.~N., 2006, mnras, 369, 655

\bibitem[{{Johnston} \& {Galloway}(1999)}]{johnston99}
{Johnston} S., {Galloway} D., 1999, mnras, 306, L50

\bibitem[{{Jones}(1993)}]{jones93}
{Jones} P.~B., 1993, mnras, 263, 619

\bibitem[{{Link}, {Epstein} \& {Lattimer}(1999){Link}, {Epstein}, \&
  {Lattimer}}]{link99}
{Link} B., {Epstein} R.~I., {Lattimer} J.~M., 1999, Physical Review Letters,
  83, 3362

\bibitem[{{Lyne} {et~al}\mbox{.}(2015){Lyne}, {Jordan}, {Graham-Smith},
  {Espinoza}, {Stappers}, \& {Weltevrede}}]{lyne15}
{Lyne} A.~G., {Jordan} C.~A., {Graham-Smith} F., {Espinoza} C.~M., {Stappers}
  B.~W., {Weltevrede} P., 2015, mnras, 446, 857

\bibitem[{{Lyne} {et~al}\mbox{.}(1996){Lyne}, {Pritchard}, {Graham-Smith}, \&
  {Camilo}}]{lyne96}
{Lyne} A.~G., {Pritchard} R.~S., {Graham-Smith} F., {Camilo} F., 1996, nat,
  381, 497

\bibitem[{{Markwardt}(2009)}]{markwardt09}
{Markwardt} C.~B., 2009, in Astronomical Society of the Pacific Conference
  Series, Vol. 411, Astronomical Data Analysis Software and Systems XVIII,
  {Bohlender} D.~A., {Durand} D., {Dowler} P., eds., p. 251

\bibitem[{{McCulloch} {et~al}\mbox{.}(1987){McCulloch}, {Klekociuk},
  {Hamilton}, \& {Royle}}]{mcCulloch87}
{McCulloch} P.~M., {Klekociuk} A.~R., {Hamilton} P.~A., {Royle} G.~W.~R., 1987,
  Australian Journal of Physics, 40, 725

\bibitem[{{Melatos}(1997)}]{melatos97}
{Melatos} A., 1997, mnras, 288, 1049

\bibitem[{{Melatos} \& {Warszawski}(2009)}]{MelatosWarszawski09}
{Melatos} A., {Warszawski} L., 2009, apj, 700, 1524

\bibitem[{{Mochizuki} \& {Izuyama}(1995)}]{mochizuki95}
{Mochizuki} Y., {Izuyama} T., 1995, apj, 440, 263

\bibitem[{{Packard}(1972)}]{packard72}
{Packard} R.~E., 1972, Physical Review Letters, 28, 1080

\bibitem[{{Palfreyman}(2016)}]{palfreyman16}
{Palfreyman} J., 2016, The Astronomer's Telegram, 9847

\bibitem[{{Radhakrishnan} \& {Manchester}(1969)}]{radhakrishnan69}
{Radhakrishnan} V., {Manchester} R.~N., 1969, nat, 222, 228

\bibitem[{{Reichley} \& {Downs}(1969)}]{reichley69}
{Reichley} P.~E., {Downs} G.~S., 1969, nat, 222, 229

\bibitem[{{{\v C}ade{\v z}} {et~al}\mbox{.}(2016){{\v C}ade{\v z}}, {Zampieri},
  {Barbieri}, {Calvani}, {Naletto}, {Barbieri}, \& {Ponikvar}}]{cades16}
{{\v C}ade{\v z}} A., {Zampieri} L., {Barbieri} C., {Calvani} M., {Naletto} G.,
  {Barbieri} M., {Ponikvar} D., 2016, aap, 587, A99

\bibitem[{{van Eysden} \& {Melatos}(2010)}]{vanEysden10}
{van Eysden} C.~A., {Melatos} A., 2010, mnras, 409, 1253

\bibitem[{{Wang} {et~al}\mbox{.}(2000){Wang}, {Manchester}, {Pace}, {Bailes},
  {Kaspi}, {Stappers}, \& {Lyne}}]{wang00}
{Wang} N., {Manchester} R.~N., {Pace} R.~T., {Bailes} M., {Kaspi} V.~M.,
  {Stappers} B.~W., {Lyne} A.~G., 2000, mnras, 317, 843

\bibitem[{{Warszawski} \& {Melatos}(2011)}]{WarszawskiMelatos11}
{Warszawski} L., {Melatos} A., 2011, mnras, 415, 1611

\bibitem[{{Yu} {et~al}\mbox{.}(2013){Yu}, {Manchester}, {Hobbs}, {Johnston},
  {Kaspi}, {Keith}, {Lyne}, {Qiao}, {Ravi}, {Sarkissian}, {Shannon}, \&
  {Xu}}]{yu13}
{Yu} M. {et~al.}, 2013, mnras, 429, 688

\end{thebibliography}
\addcontentsline{toc}{chapter}{\textsc{bibliography}}

\end{document}